\documentclass[journal]{IEEEtran}

\usepackage{bibentry}  
\usepackage{xurl}
\usepackage{xcolor,soul,framed} %,caption

\colorlet{shadecolor}{yellow}
\usepackage[pdftex]{graphicx}
\graphicspath{{../pdf/}{../jpeg/}}
\DeclareGraphicsExtensions{.pdf,.jpeg,.png}

\usepackage[cmex10]{amsmath}
\usepackage{array}

\usepackage{eqparbox}
\usepackage{url}
\usepackage{multirow}
\usepackage{tabularx}

\usepackage{booktabs}
\usepackage{dingbat}
\usepackage{diagbox}
\newcolumntype{C}{>{\centering\arraybackslash}X}
\usepackage{adjustbox}
\usepackage{pifont}
\usepackage{mdwmath}
\usepackage{mdwtab}

\usepackage{tikz}

\newcommand\copyrighttext{%
  \footnotesize ©2021 IEEE.  Personal use of this material is permitted.  Permission from IEEE must be obtained for all other uses, in any current or future media, including reprinting/republishing this material for advertising or promotional purposes, creating new collective works, for resale or redistribution to servers or lists, or reuse of any copyrighted component of this work in other works. }
\newcommand\copyrightnotice{%
\begin{tikzpicture}[remember picture,overlay]
\node[anchor=south,yshift=7pt] at (current page.south) {\fbox{\parbox{\dimexpr\textwidth-\fboxsep-\fboxrule\relax}{\copyrighttext}}};
\end{tikzpicture}%
}

\begin{document}
%COMMON THEMES FROM REVIEWERS COMMENTS

\bstctlcite{IEEEexample:BSTcontrol}
    \title{Explanations in Autonomous Driving: A Survey}
  \author{Daniel Omeiza,~\IEEEmembership{}
      Helena Webb,
      Marina Jirotka,
      Lars Kunze~\IEEEmembership{}
     % <-this % stops a space

  \thanks{
  This work was supported by the UK's Engineering and Physical Sciences Research Council (EPSRC) through the project RoboTIPS: Developing Responsible Robots for the Digital Economy, grant reference EP/S005099/1. It was also supported by the Assuring Autonomy International Programme (Demonstrator project: Sense-Assess-eXplain (SAX)), a partnership between Lloyd's Register Foundation and the University of York; the Trustworthy Autonomous Systems Hub's project RoAD (Responsible AV Data), grant reference EP/V00784X/1; and the Responsible Technology Institute, Oxford.}
  \thanks{Daniel Omeiza and Marina Jirotka are with the Department of Computer Science, University of Oxford, UK (e-mail: daniel.omeiza@cs.ox.ac.uk).}% <-this % stops a space
\thanks{Helena Webb is with the School of Computer Science, University
of Nottingham, Nottingham, UK (e-mail: helena.webb@nottingham.ac.uk)}% 

  \thanks{Lars Kunze is with the Oxford Robotics Institute, Department of Engineering Science, University of Oxford, UK (e-mail: lars@robots.ox.ac.uk)}% 
  }

% The paper headers
\markboth{
}{Roberg \MakeLowercase{\textit{et al.}}: High-Efficiency Diode and Transistor Rectifiers}

% ====================================================================
\maketitle
% === ABSTRACT ====================================================================
% =================================================================================
\begin{abstract}

The automotive industry has witnessed an increasing level of development in the past decades; from manufacturing manually operated vehicles to manufacturing vehicles with a high level of automation. With the recent developments in Artificial Intelligence (AI), automotive companies now employ blackbox AI models to enable vehicles to perceive their environments and make driving decisions with little or no input from a human. With the hope to deploy autonomous vehicles (AV) on a commercial scale, the acceptance of AV by society becomes paramount and may largely depend on their degree of transparency, trustworthiness, and compliance with regulations. The assessment of the compliance of AVs to these acceptance requirements can be facilitated through the provision of explanations for AVs' behaviour.
Explainability is therefore seen as an important requirement for AVs. AVs should be able to explain what they have ‘seen’, done, and might do in environments in which they operate.
In this paper, we provide a comprehensive survey of the existing body of work around explainable autonomous driving. First, we open with a motivation for explanations by highlighting and emphasising the importance of transparency, accountability, and trust in AVs; and examining existing regulations and standards related to AVs. Second, we identify and categorise the different stakeholders involved in the development, use, and regulation of AVs and elicit their explanation requirements for AV. Third, we provide a rigorous review of previous work on explanations for the different AV operations (i.e., perception, localisation, planning, control, and system management). Finally, we identify pertinent challenges and provide recommendations, such as a conceptual framework for AV explainability. This survey aims to provide the fundamental knowledge required of researchers who are interested in explainability in AVs.
\end{abstract}
\makeatletter
%%%%%%%%%%%%%%%%%%%%%%%%%%%%%% User specified LaTeX commands.
\copyrightnotice

% === KEYWORDS ====================================================================
% =================================================================================

%\copyrightnotice

% For peer review papers, you can put extra information on the cover
% page as needed:
% \ifCLASSOPTIONpeerreview
% \begin{center} \bfseries EDICS Category: 3-BBND \end{center}
% \fi
%
% For peerreview papers, this IEEEtran command inserts a page break and
% creates the second title. It will be ignored for other modes.
\IEEEpeerreviewmaketitle

% ====================================================================
% ====================================================================
% ====================================================================

% === I. INTRODUCTION =============================================================
% =================================================================================
\section{Introduction}

\IEEEPARstart{T}{he} advent of autonomous vehicles (AVs) is a significant milestone in the automotive industry. The increasing growth rate in the industry is considered to be a result of the increasing research knowledge in vehicle dynamics. Moreover, the enhancements of sensing devices (e.g., LiDAR \cite{lidar1, lidar2} and Radar \cite{radar1, radar2}) and the emergence of deep learning algorithms have also contributed to the industry growth. %REWORD
Despite the technological advancements, the successful deployment of AVs in the real world may greatly depend on users' perception of safety, and in turn, trust. Accidents caused by AV technologies~ \cite{tesladeaths,model1,teslacrash,ekim7,ekim8,ekim9} hamper trust~\cite{yurtsever2020survey}. Thus, effective ways to build confidence and trust in AVs need to be investigated. Explanation provision is considered as a way of building confidence and trust in AV technologies~\cite{ha2020effects,koo2015did}.

People (e.g., developers and regulators) can benefit from explanations that explain the actions and behaviour of autonomous systems. Consider the Molly problem\footnote{https://www.itu.int/en/ITU-T/focusgroups/ai4ad/Pages/MollyProblem.aspx} described by the International Telecommunication Union (ITU) in relation to AVs. A young girl called Molly was crossing the road alone and was hit by an unoccupied self-driving vehicle. There were no eye-witnesses. Post-hoc explanations containing the vehicle’s observations, the road rules, and traffic signs it acted on will serve as clues to the causes of the accident. These clues will inform the accident investigation process. Moreover, system auditors can also benefit from an easier auditing process in the presence of explanations to provide an assurance of safety.

Further, the general stress on explainable AI (XAI) and the ``right to explanation” as stipulated in the General Data Protection Regulation (GDPR) \cite{gdpr} underscores the essence of explanations in complex systems, especially when they are powered by black-box models. Such complex systems, such as, autonomous vehicles are likely to make decisions that are strange and confusing to end-users. This is because AI decisionality is inherently different from human decision-making processes~\cite{cunneen2019autonomous}. This is especially the case for vehicles with a high level of automation, e.g., vehicles in \textit{SAE} Level 3 or above---the Society of Automotive Engineers (SAE) automation categorisation tends to be the predominant categorisation model used by automotive engineers. This categorisation model classifies vehicles into five levels based on automation~\cite{sae2016taxonomy}. As highly automated vehicles make high-stake decisions that can significantly affect end-users, the vehicles should explain or justify their decisions to meet set transparency guidelines.

A large body of literature on explainable AI focuses on explaining single neurons or a single artificial neural network model while only a handful focuses on explaining an entire goal-based system like autonomous vehicles which possess unique architecture and various interacting sub-systems. Providing explanations for the behaviour of such goal-based systems is therefore essential.

In this paper, we provide a structured and comprehensive overview of the recent work on explanations in autonomous driving. Previous literature surveys such as~\cite{peekingbb, bbeval, samek2016interpreting} focused on approaches aimed at `opening' black-box machine learning mechanisms (data-driven XAI) applied in deep learning. In contrast, Anjomshoae et al.~\cite{sule} provided a systematic literature review generally on explainable agencies (i.e., explaining the behaviour of goal-driven agents and robots) which entailed the use of descriptive statistics to show the amount of research done around explainable agencies with no particular focus on autonomous driving. Also, Zablocki~et al.'s~\cite{zablocki2021explainability} provided a survey that focused on vision-based autonomous driving systems.

Our paper aims to fill the gap in the academic literature by providing a comprehensive survey on explanations for the behaviour of AVs at different aspects of operations (i.e., perception, localisation, planning, vehicle control, and system management) with the requirements of different stakeholders in mind. We also propose recommendations such as a conceptual framework for AV explainability including possible user interfaces, and some regulatory considerations for AV explainability. In this paper, a stakeholder is an individual or an agent whose roles involve interaction with an AV. While there may be some technology overlap between autonomous vehicles, uninhabited aerial vehicles (UAV) and autonomous underwater vehicles (AUV), we will focus on autonomous vehicles in this survey to enable us to cover enough depth. 

We note that there are different definitions of explanations in psychology~\cite{dodwell1960causes}, philosophy~\cite{zalta1995stanford} and AI~\cite{ciatto2020agent, omicini2020not}; we assume a more general meaning by referring to an explanation as a piece of information presented in an intelligible way as a reason or part of a reason for an outcome, event or an effect in text, speech, or visual forms. We refer to explainability as the ability of a system to support the provision of this form of explanations. 
We refer to interpretable techniques as techniques that are transparent enough to support meaningful interpretations to their intended audience (mostly developers). We do not expect laypeople to be able to easily and quickly make sense of interpretable models~\cite{poursabzi2021manipulating}. However, intelligible explanations (e.g., in natural language) for the outputs of the interpretable models will be beneficial to laypeople. This type of explanations are easier to obtain from interpretable models as shown in~\cite{nahata2021,stepin2021factual}.

The rest of this paper is organised into 9 sections: Section~\ref{sec:sec2} presents the general need for explanations in autonomous vehicles. Section~\ref{sec:sec3} presents and discusses the regulations and standards related to explanations in AVs. The different stakeholders who interact with AVs are identified and categorised in Section~\ref{sec:stakeholders}. The categorisation system defined in Section~\ref{sec:stakeholders} is used in the rest of the paper. Section~\ref{sec:dim} broadly categorises explanations into many dimensions and provides several literature references for the different categories. Section~\ref{sec:expops} describes the core operations of an AV and reviews existing work on explanations in relation to the different core AV operations. These operations include perception, localisation, planning, and control. Section~\ref{sec:hmi} examines AV system management. System management involves event data recorders and human-machine interaction which are crucial in explanations. Section~\ref{sec:challenges} discusses challenges in the explainable AV landscape and provides recommendations for future work in the field. Section~\ref{sec:conclude} concludes the paper with a summary.

\section{Need for Explanations}
\label{sec:sec2}
The need for explanations in autonomous vehicles stems from the increasing concerns for transparency and accountability of autonomous vehicles. It is believed that explanations are one way of achieving these goals. In this section, we discuss the need for explanation in the light of transparency, accountability, and trust.

\subsection{Transparency and Accountability}
One generally agreed upon notion of accountability is associated with the process of being called `to account’ to some authority for one’s actions~\cite{jones1992search}. Accountability, in broad terms, often encompasses closely related concepts, such as responsibility and liability~\cite{martinho2021ethical}. Mulgan~\cite{mulgan2000accountability} elucidated that accountability entails responsibility but, unlike responsibility, it requires explanations about actions and it cannot be shared. Meanwhile, liability is a legal or financial responsibility~\cite{collingwood2017privacy}. In the human and machine context, Doshi-Velez et al.~\cite{doshi2017accountability} conceptualise accountability as the ability to determine whether the decision of a system was made in compliance with procedural and substantive standards, and importantly, to hold one responsible when there is a failure to meet the standards. In autonomous driving, accountability becomes a challenging issue mainly because of the various operations involved (e.g., perception, planning, controls,  system management among others) that demand inputs from multiple stakeholders; this can result in responsibility gaps.

As identified by Mulgan~\cite{mulgan2000accountability}, achieving accountability requires social interaction and exchange. At one end, the requester of an account seeks answers and rectification while at the other end, the respondent or explainer responds and accepts responsibility if necessary. In the context of this paper, the AV is being called by a stakeholder to provide an account; we expect the AV to provide an account in the form of an explanation that is intelligible to the requester to facilitate the assignment of responsibilities. There have been debates on how responsibility should be allocated for certain AV accidents. Companies have stated the need for clear rules to be set in advance. For example, Honda has mentioned that it is necessary to put legal frameworks in place in order to clarify where the responsibility lies in case of the occurrence of an accident after the realisation of fully automated driving~\cite{honda2015}. Technical solutions are also being put forward. One such example is the proposal for the use of a `blackbox', similar to a flight recorder in an aircraft, to facilitate investigations~\cite{bmw2016}. Shashua and Shalev-Shwartz~\cite{shashua2017plan} also advocated for the use of mathematical models to clarify faults in order to facilitate a conclusive determination of responsibility. 

The social aspect of accountability described by Mulgan~\cite{mulgan2000accountability}, will demand that the aforementioned recommended approaches are able to plug into explanation mechanisms where causes and effects of actions can be communicated to the relevant stakeholders in intelligible ways. In addition to accountability for accident cases, which has gained much attention in the industry reports, actions resulting in undesired, discriminatory, and inequitable outcomes also need to be accounted for. This means that stakeholders such as passengers or auxiliary drivers who may not have direct involvement in the management of the AVs should be able to instantaneously request accounts as intelligible explanations for such undesired actions when they occur.

\subsection{Trust}
Faas et al.~\cite{m2021calibrating} argued that research investigating trust in automation has been around for decades, i.e., since the introduction of interpersonal trust theories into the human-machine interaction domain by~\cite{muir1987trust, muir1994trust, lee1992trust, m2021calibrating}. While various definitions of trust in automation have been proposed, the most commonly adopted definition is that put forward in~\cite{lee2004trust}. Lee and See~\cite{lee2004trust} consider trust as a social psychological concept that is important for understanding automation partnership. The authors emphasise that trust is the attitude that an agent or automation will help an individual to achieve their goals in a situation characterised by uncertainty and vulnerability. Trust in automation, as made evident in~\cite{biros2004influence, hergeth2016keep, muir1996trust}, has significantly influenced the acceptance of and reliance on automated systems. As opposed to a binary categorisation, trust can be more finely calibrated so that an individual's trust levels on an automated system adequately reflects the actual capabilities and functional scope of an automated system. This trust calibration is considered to be an important requirement for safe and efficient human-machine interaction~\cite{kraus2020more, muir1987trust}. While calibration is useful, miscalibrated trust is disastrous as it can lead to distrust or overtrust (i.e., excessive trust). This will either make the user under-use the system or use the system beyond the scope of its functionalities~\cite{m2021calibrating}. The process of using available information to assess and learn about the trustworthiness of a system to adapt trust levels is referred to as trust calibration~\cite{khastgir2018calibrating, kraus2020more, kraus2020psychological}.

Information about the functioning modes of an AV at the user's disposal can help the user create a better understanding of the AV's behaviour, eventually adding to the user's knowledge base~\cite{hoffman2018metrics}, and helpful for constructing calibrated trust. This information could be presented as explanations of the operational modes and behaviour of a complex system, such as an AV, especially when it acts outside the expectations of the user. We note that trust can break down when there are frequent failures without adequate explanations, and regaining trust once lost can be challenging~\cite{dzindolet2003role, madhavan2007effects}. For example, previous reports on AV accidents may have a negative impact on calibrated trust in AVs. According to Hussainet al.~\cite{hussain2018autonomous}, a serious challenge evident in intelligent transport systems is the lack of trust from the consumer's perspective. The public fears that the claims on accidents reduction through the introduction of AVs may be misleading as they consider human drivers to be better than AVs in handling~\cite{hussain2018autonomous} unforeseen and uncharacteristic traffic  situations. Abraham et al.~\cite{abraham2016autonomous} also reported that the consumers' perception of trust is still not as high as expected in spite of the great potentials promised by AVs, claiming that the public is still hesitant about the technology, and still feel uncomfortable using it. Trust is therefore imperative for achieving widespread deployment and use of AVs.

Researchers (e.g., in  \cite{hoffman2017explaining}) suggest that the provision of meaningful explanations from AVs to stakeholders (e.g., passengers, pedestrians and other road participants) is one way to build the necessary trust in AV technology. Other empirical studies~\cite{koo2015did, ha2020effects, omeiza2021not} have shown that the provision of explanations in AVs can influence trust. While it has been argued (e.g., in~\cite{hergeth2016keep, payre2016fully, rajaonah2006trust}) that trust is a substantial subjective predicting factor for the adoption of automated driving systems, several studies have shown the importance of viewing trust formation and calibration in AVs as a temporal process influenced by prior information or background knowledge~\cite{beggiato2013evolution, kraus2019two}.
Explanation provision in autonomous driving over time is therefore crucial. In the following section, we will discuss explanations from the regulatory perspective.

\section{Regulations, Standards, and Stakeholders}
\label{sec:sec3}
\subsection{Explanation and AV Regulations}
There are increasing concerns about the collection and use of personal data in algorithms that make critical decisions about people in domains like healthcare, finance, insurance, and criminal justice. We explore this section mainly from the GDPR perspective. The European Union GDPR implemented in 2018 aims to provide more control rights to individuals over their personal data~\cite{turing}.

The GDPR also sets guidelines related to the explanation of decisions made based on users' data. The GDPR guideline mandates that controllers (entities handling people’s personal data) provide meaningful information about the logic involved in the decisions made based on people's data and what the likely consequences are for individuals. It also demands the appropriate use of mathematical or statistical procedures on such data. This is commonly referred to as the `right to explanation’. In addition, the GDPR Article 12 on transparency demands that the provision of information/explanation to data subjects must be done in an intelligible way (i.e., in a clear and easy to understand manner~\cite{lim2009and}). These clauses highlight the user's right to question the decision of a system and the demand for explanations, especially when decisions are made based on their data. 

An autonomous vehicle can potentially be used to collect sensitive information from users either legally or illegally. By tracking an AV, a passenger's location is known, a passenger's frequent routes can be determined, as can the time of the day they typically travel. 
% These and many more information can be used to personalise passengers' in-vehicle experience~\cite{collingwood2017privacy}. 
Hence, autonomous vehicles should not be exempt from the GDPR clause on explanations, especially when they operate in the regions where this regulation holds. Consequently, the acceptance and adoption of AVs will introduce new regulations and standards challenges for governments and experts~\cite{hussain2018autonomous}. Many current regulations which relate to human drivers will need to be revised when a vehicle drives itself in society~\cite{khan2012policy}. That said, a few guidelines have been developed recently to govern autonomous driving. Closely related to AV explainability, the National Transportation Safety Board (NTSB) made a call for efficient event data recording in AVs which could facilitate the provision of plausible and faithful explanations to ensure correct accident investigation \cite{teslacrash}. More details on event data recording are provided in Section~\ref{sec:hmi}. Countries are also defining new regulations for autonomous vehicles, for example, the Scottish Law Commission's regulation for autonomous vehicles~\cite{scottland}. Unfortunately, these regulations(~\cite{scottland,dentons}) do not directly address the explainability challenge. We further elaborate on this in Section~\ref{sec:challenges}.

Refer to~\cite{apexai} for further details on regulations and ethical guidelines. 

\subsection{AV Standards}
Intelligent Transport Systems (ITS) apply advanced electronics, information and communications technologies to roads and automobiles. This is done to collect, store, and provide traffic information in real-time for convenient and safe transport; improved reliability, efficiency, and quality; and the reduction in energy consumption~\cite{standards}.
The international standard organisation technical committee 204 (ISO TC204), the IEEE, and related standard organisations have set standards for AVs and ITS in general. The IEEE Initiatives, in particular, has a vision for prioritising human well-being with autonomous and intelligent systems, and the assessment of gaps in standardisation for safe autonomous driving. These standards directly or indirectly demonstrate the necessity of explainability in AVs. In Table~\ref{tab:stds}, we identified standards that are related to safety and information/explanation provision in AVs. We categorised these standards into two sets: Human safety-related standards and information or data exchange related standards. For more details on AV related standards, refer to the ISO report on intelligent transport systems in~\cite{standard3} and the Apex.AI document on automated mobility~\cite{apexai}. To the best of our knowledge, there is currently no explainability standards for autonomous driving. To address this gap, we have elaborated in Section~\ref{sec:challenges} on how some of the existing AV regulations and the related standards highlighted in Table~\ref{tab:stds} can be extended to directly support explainability in autonomous driving. 

We now describe different stakeholders involved in AV explanations in the next section.

\begin{table*}[]
\begin{center}
\caption{Selected standards for autonomous vehicles. These standards underline the importance of safe, transparent, and explainable AVs.}
\label{tab:stds}
\begin{tabular}{|c|l|l|}
\cline{1-3}
\multicolumn{1}{|c|}{\textbf{Aim}} & 
  \multicolumn{1}{c|}{\textbf{Standard \& Description}} &
  \multicolumn{1}{l|}{\textbf{ Stakeholder}} \\\hline
 
  &
 \multicolumn{1}{m{10cm}|}{\textbf{ISO 19237:2017} Pedestrian detection and collision mitigation systems}
&  \\ \cline{2}

 &
\multicolumn{1}{m{10cm}|}{ \textbf{ISO 22078:2020}  Bicyclist detection and collision mitigation systems}

&  \\ \cline{2}

 &
  \multicolumn{1}{m{10cm}|}{\textbf{ISO 26262:2011: Road vehicles – Functional safety.} An international standard for functional safety of electrical and/or   electronic (E/E) systems in production automobiles (2011). It addresses possible hazards caused by the malfunctioning behaviour of E/E safety-related systems, including the interaction of these systems.}  & \\  \cline{2}

   &
  \multicolumn{1}{m{10cm}|}{\textbf{ISO 21448:2019: Safety Of The Intended Functionality (SOTIF).} Provides guidance on design, verification and validation   measures. Guidelines on data collection (e.g. time of day, vehicle speed,  weather conditions) (2019). (complementary to ISO 26262).} & \multirow{7}{*}[0pt]{\parbox{2cm}{\textbf{Class B and C}\\ AV Developers,\\ Regulators,\\System Auditors,\\ Accident Investigators,\\ Insurer}} 
  
  \\ \cline{2} 
 &
  
  \multicolumn{1}{m{10cm}|}{\textbf{UL 4600: Standard for Safety for Evaluation of Autonomous Products.} a safety case approach to ensuring autonomous product safety in general, and self-driving cars in particular.}  & \\ \cline{2}
 
   \multirow{0}{*}[1pt]{\parbox{2cm}{\textbf{Human Safety}}} 
  &
  
    \multicolumn{1}{m{10cm}|}{\textbf{SaFAD: Safety First for Automated Driving.}  White paper by eleven companies from the automotive industry and automated driving sector about frameworks for development, testing and validation of safe automated passenger vehicles (SAE Level 3/4).} & \\ \cline{2}

  &
 
   \multicolumn{1}{m{10cm}|}{\textbf{RSS (Intel) / SFF (NVIDIA): Formal Models \& Methods} to evaluate safety of AV on top of ISO 26262 and ISO 21448 (proposed by   companies).} & \\ \cline{2} 
 
   &
   
     \multicolumn{1}{m{10cm}|}{\textbf{IEEE Initiatives:} ``Reliable,   Safe, Secure, and Time-Deterministic Intelligent Systems (2019)''; ``A Vision for Prioritizing Human Well-being with Autonomous and Intelligent Systems''  (2019); ``Assessment of standardization gaps for safe autonomous driving (2019)''.} & \\ \cline{2} 
 &
   \multicolumn{1}{m{10cm}|}{\textbf{The Autonomous:} Global safety reference, created by the community leading automotive industry players, which facilitates the adoption of autonomous mobility on a grand scale (2019).}  & \\ \hline 
 
 &
  \multicolumn{1}{m{10cm}|}{\textbf{ISO/TR 21707:2008: Integrated transport information, management, and control---Data quality in intelligent transport systems (ITS)}. ``specifies a set of standard terminology for defining the quality of data being exchanged between data suppliers and data consumers in the ITS domain'' (2018).} & \multirow{7}{*}[0pt]{{\parbox{2cm}{\textbf{Class A and C}\\ Passengers, Auxiliary Drivers, \\ Pedestrians, Regulators, System Auditors,\\ Accident Investigators\\ Insurers}}}
 
\\ \cline{2}
 &
  \multicolumn{1}{m{10cm}|}{\textbf{ISO 13111-1:2017: The use of personal ITS station to support ITS service provision for travellers.} ``Defines the general information and use cases of the applications based on the personal   ITS station to provide and maintain ITS services to travellers including drivers, passengers, and pedestrians'' (2017).} & \\ \cline{2}
 
 \multirow{0}{*}{\parbox{2cm}{\textbf{Information/Data Exchange}}}
 &
  
  \multicolumn{1}{m{10cm}|}{\textbf{ISO 15075:2003: In-vehicle navigation systems---Communications message set requirements.} ``Specifies message content and format utilized by in-vehicle navigation systems'' (2003).}  &\\ \cline{2} 
 &
  
\multicolumn{1}{m{10cm}|}{\textbf{ISO/TR 20545:2017: Vehicle/roadway warning and control systems.} ``Provides   the results of consideration on potential areas and items of standardization   for automated driving systems'' (2017).}  & 
\\ \cline{2}
 &
 \multicolumn{1}{m{10cm}|}{\textbf{ISO 17361:2017:} Lane departure warning.}  & \\ \cline{2}

&
\multicolumn{1}{m{10cm}|}{\textbf{ISO/DIS 23150:} Data communication between sensors and data fusion unit for automated driving functions.}

& \\ \hline
  
\end{tabular}
\end{center}
\end{table*}

\subsection{Stakeholders}
\label{sec:stakeholders}
Explanation provision in autonomous driving has many personas due to the different purposes of explanations. The level of detail (in terms of information) anticipated by the explanation recipients, the explanation type and the mode of communication vary with respect to the type of recipient and purpose for the explanation. This highlights the importance of explanation personalisation with respect to stakeholders. Personalisation is seen to be crucial for the generation of intelligible or understandable explanations~\cite{meske2020explainable, kouki2019personalized, shin2021effects}. While lay users who lack technical domain expertise may be satisfied with a user-friendly explanation that requires less background knowledge to interpret, developers and engineers would prefer a finely detailed explanation with technical terms that would support a deeper conception of the internal functioning of a model \cite{xaifolk}. In this light, the consideration of the persona of the explainee is necessary \cite{xaivarieties}. Going forward, we refer to anyone who has to engage with an explanation as a stakeholder. Having identified the typical personas in the literature, we divided stakeholders into three broad categories: Class A (all types of end-users and society), Class B (all technical groups, e.g. developers), and Class C (all forms of regulatory bodies including insurers). See the further description: 
\begin{enumerate}
    \item \textbf{Class A: End-Users}
    \begin{itemize}
        \item Passenger: this is the in-vehicle agent who may interact with the explanation agency in the AV but is not responsible for any driving operation.
        \item Auxiliary Driver: This is a special in-vehicle passenger who may also interact with the explanation agency in the AV and can also participate in the driving operations. This kind of participant may mainly exist in SAE level 3 and 4 vehicles.
        \item Pedestrian: this is the agent outside the AV (external agent) who may interact with the AV to convey intentions either through gestures or an external human-machine interface (eHMI).
        \item Pedestrian with Reduced Mobility (PRM): this is the agent outside the AV (external agent) who may interact with the AV to convey intentions either through gestures or an external human-machine interface (eHMI) but have reduced mobility capacity (e.g., pedestrian in a wheelchair).
        \item Other Road Participants: these are other agents outside the AV (external agent) who may interact with the AV to convey intentions either through gestures or an external human-machine interface (eHMI) (e.g., cyclists, other vehicles).
        
    \end{itemize}
    \item \textbf{Class B: Developers and Technicians}
    \begin{itemize}
        \item AV Developer: the agent who develops the automation software and tools for AVs.
        \item Automobile Technicians: the agent who repairs and maintains AVs. 
    \end{itemize}
    \item \textbf{Class C: Regulators and Insurers}
    \begin{itemize}
        \item System Auditor: the agent who inspects AV design processes and operations in order to ascertain compliance with regulations and guidelines.
        \item Regulator: the agent who sets guidelines and regulations for the design, use, and maintenance of AVs.
        \item Accident Investigator: the agent who investigates the cause of an accident in which an AV was involved.
        \item Insurer: the agent who insures the AV against vandalism, damage, theft, and accidents.
    \end{itemize}
\end{enumerate}

In the next section, we provide a categorisation of explanations based on methodologies, and situate the different stakeholders in the categorisation.

\section{Explanation Categorisations}
\label{sec:dim}
Explanations serve different functions in different contexts~\cite{xaifolk}. Therefore, the methods of generation and evaluation are context and purpose-dependent~\cite{binns2018s}. Wang et al.~\cite{wang2019designing} identified three approaches that have been adopted in the academic literature in either developing or evaluating explanations.

First, the authors in~\cite{wang2019designing} highlighted the existence of \textbf{unvalidated guidelines} for the design and evaluations of explanations. They claim that these kinds of guidelines are based on authors' experiences with no further substantial justification. Hence, explanation generation algorithms that generate explanations as short rules~\cite{lakkaraju2016interpretable}, or those that apply influence scores \cite{bussone2015role} (such as partial dependence plots~\cite{caruana2015intelligible}) without sufficient justification for the explanation choices made are assumed to be based on unvalidated guidelines. Thus, the explanations generated by these algorithms may not be appropriate for class A stakeholders due to the low intelligibility quality of the explanations~\cite{xaifolk}.

Second, researchers suggested (in \cite{zhu2018explainable}) that understanding users' requirements might be helpful in explainable AI research. It is on this premise that some research on explanation design approaches have been thought to be \textbf{empirically derived}. This type of research elicits explanation requirements from user surveys in order to determine the right explanation for a use-case with explanation interfaces~\cite{wang2019designing}. For instance, explanation frameworks have been proposed for recommender systems~\cite{herlocker2000explaining}, case-based reasoning~\cite{roth2004explanations}, intelligent decision aids~\cite{silveira2001semiotic}, and intelligible context-aware systems~\cite{lim2009assessing} upon the elicitation of users’ requirements through surveys and user studies. Through user studies, Lim and Anind~\cite{lim2009assessing} examined explanations based on intelligibility types. The intelligibility types used were: `why’ (factual), `why not’ (contrastive), `what if’ and `how to’ (counterfactual) explanations which are considered relevant for filtering causes for an effect. We interchangeably refer to these intelligibility types as causal filters or investigatory queries in this paper.

Third, some explanation design methods are derived from \textbf{psychological constructs from formal theories} in the academic literature \cite{wang2019designing}. Some of these methods (e.g., in \cite{hoffman2017explaining}) draw on philosophy, cognitive psychology, social science, and AI theories to inform explanation design for explanation frameworks. For example, Akula et al.~\cite{akula2019x} employed the Theory of Mind (ToM) in the development of an explanation framework (X-ToM). The authors in~\cite{akula2019x} claimed that in their explanation framework, the mental representations in ToM were incorporated to learn an optimal explanation policy that took into account human's perception and beliefs. Simply put, a policy, as used in this context is an agent's strategy for achieving a goal~\cite{sutton2018reinforcement}. Theory of mind involves explaining people’s
behaviour on the basis of their
minds: their knowledge, their
beliefs, and their desires~\cite{frith2005theory}. We note that there are criticisms of the theory of mind and mental models. However, this is out of the scope of this paper.

We use the three discussed methodologies as one of our categorisation dimensions. Explanations methods that are mainly based on the researcher's experience without further user studies to justify claims are categorised under unvalidated guidelines (UG). Those that adopted a user study to elicit users experience are categorised as empirically derived (ED), and those that built on psychology theories as are categorised under psychological constructs from formal theories (PC). Other dimensions for our categorisation include causal filter, explanation style, interactivity, dependence, system, scope, stakeholders, and operation.

The description of the various dimensions of explanation is detailed below.
\begin{table}[]
\caption{Causal filters and example investigatory queries.}
\label{tab:extypes}
\begin{tabular}{lll}
\cline{1-3}
\textbf{Causal Filter} & \textbf{Class}  & \textbf{Example Query }     \\ \cline{1-3}
Why Not (Contrastive) & Causal & why did you not do Y?  \\ 
Why (Factual)   & Causal  & why did you do X?      \\ 
What If (Counterfactual)    & Causal & what would you do if Z? \\ 
What & Non-Causal & what are you doing? \\ \cline{1-3}
\end{tabular}

\end{table}
\paragraph {Causal Filters} explanations resulting from causal filters use selected \emph{causes} relevant to interpreting an observation, with respect to existing knowledge \cite{wang2019designing}. The explanations provided in this category are assumed to be usually generated by causal filters or investigatory queries like \emph{why, why not, how to,} and \emph{what if} \cite{lim2009and}. These causal filters are assumed to produce  explanations that could be factual (e.g. `why' explanation), contrastive (`why not’ explanation), or counterfactual (`how to' and `what if' explanation). See Table~\ref{tab:extypes}.
\paragraph{Explanation Style} explanations are categorised based on the type of information or elements referenced in the explanation and the forms they are presented in~\cite{binns2018s}.
\begin{itemize}
    \item {Input Influence:} a list of input variables is presented along with quantitative measures of their influence (either positive or negative) on a decision.
    \item{Sensitivity:} shows what magnitude of change is required in an input variable in order to change the output class. Note that this is different from sensitivity used in machine learning evaluation.
	\item {Case-based:} picks out a relevant case from the model’s training data that is most similar to the decision made, which is then used to explain.
    \item{Demographic:} explanation provides aggregate statistics of previous outcomes for people with the same demographics.
\end{itemize}

\paragraph{Model Dependence} in this context refers to the possibility of having an explanation method that can be used to explain any type of autonomous driving model (e.g., perception models and motion planning models). If the possibility exists, the explanation method is considered to be \textit{model agnostic}. Otherwise, it is regarded as \textit{model specific}. Two popular model-agnostic explanation techniques are SHAP~\cite{lundberg2017unified} and LIME~\cite{ribeiro2016should}. Although LIME and SHAP explanation techniques can be useful in autonomous driving, to the best of our knowledge, only SHAP has been used in the context of autonomous driving~\cite{nahata2021}.
\paragraph{Interactivity} this refers to the possibility of a stakeholder raising follow-up questions as a way of demanding further explanations. The conversational style of explaining~\cite{miller} allows for this.
\paragraph{System Type} this refers to the nature of the system that the explanation technique is primarily designed for. It could be an explanation technique for \textit{data-driven} systems (e.g., explaining the output of a machine learning model) or a \textit{goal-driven} system (e.g., explaining the behaviour of an autonomous agent based on plans and goals)~\cite{sule}. In more detail, an explanation method that explains a deep learning model trained on driving scene images or video is data-driven while one that explains plans (or change in plans) in the absence of a trained machine learning model is referred to as goal-driven in this context. 

\paragraph{Scope} in this context refers to the coverage of the explanation in terms of the system's parts. We adapt terminologies from the explainable AI (XAI) domain. While global explanation explains a model’s decision-making process in general, a local explanation explains a single prediction in XAI \cite{lundberg2020local,van2019global}. The term global explanation in this paper is used to refer to an explanation that explains the entire behaviour of the AV. In contrast, a local explanation refers to an explanation that only explains a subset of the AV's behaviour. Nahata et al.~\cite{nahata2021} proposed a tree explanation technique that can provide both factual (\textit{why}) and counterfactual (\textit{what if}) explanations for an AV collision risk model. Users can specify simple constraints for generating counterfactual explanations (e.g., setting the desired counterfactual output to be explained).

A representative subset of previous works where an explanation technique was primarily discussed or implemented in the context of autonomous driving are shown in Table~\ref{tab:papers}. Note that \cite{chakraborti2017plan,raman2013sorry} are mainly on robot plan explanations but are applicable to autonomous vehicles. While attention maps are commonly regarded as explanations in the machine learning literature, Jain and Wallace~\cite{jain2019attention} argued against this notion by claiming, based on the outputs of experiments, that attentions maps are not explanations. Consequently, Wiegreff and Pinter~\cite{wiegreffe2019attention} disproved this claim and argued that such a claim depends on one’s definition of explanation and that prior work against the effectiveness of attention maps for explanations does not disprove the usefulness of attention mechanisms for explainability. We agree that attention maps and heatmaps are not effective in some cases but are however useful. Therefore, we include relevant works on attention maps and heatmaps in this survey. See Table~\ref{tab:papers} for an overview.

\newcommand{\cmark}{\text{\ding{51}}}
\newcommand{\xmark }{\text{\ding{55}}}
\begin{table*}[]
\caption{Summary of explanations categories. The table includes a subset of the reviewed papers where each or a subset of the explanation categories was mentioned in the context of autonomous embodied agents.\\ \textbf{Stakeholders:} Class A---Passenger (PA), Pedestrian (PE), Pedestrian with Reduced Mobility (PRM), Other Road Participants (ORP), Auxiliary Driver (AD). Class B---Developer (DV), Auto-Mechanic (AM). Class C---System Auditor (SA), Regulator (RG), Insurer (IN), Accident Investigator (AI).\\
\textbf{Methods:} Unvalidated Guidelines (UG), Empirically Derived (ED), Psychological Constructs from Formal Theories  (PC). \\
\textbf{Operations:} Perception (P), Localisation (L), Planning (PL), Control (C), System Management (M)}
\begin{tabular}{|l|l|l|l|l|l|l|l|l|l|l|l|l|l|l|l|l|l|l|}
\hline
\multirow{2}{*}[0pt]{\textbf{References}} &
  \multicolumn{3}{l|}{\textbf{Causal Filter}} &
  \multicolumn{4}{l|}{\textbf{Explanation Style}} &
  \multicolumn{2}{l|}{\textbf{Interactivity}} &
  \multicolumn{2}{l|}{\textbf{Dependence}} &
  \multicolumn{2}{l|}{\textbf{System}} &
  \multicolumn{2}{l|}{\textbf{Scope}} &
  \multirow{2}{*}[0pt]{\rotatebox[origin=c]{90}{\parbox{2cm}{\textit{Method}}}} &
  \multirow{2}{*}[0pt]{\rotatebox[origin=c]{90}{\parbox{2cm}{\textit{Stakeholders\\(Class)}}}} &
  \multirow{2}{*}[0pt]{\rotatebox[origin=c]{90}{\parbox{2cm}{\textit{Operation}}}}
  \\ \cline{2-16}
 &
  \rotatebox[origin=c]{90}{\textit{Factual}} &
   \rotatebox[origin=c]{90}{\textit{Contrastive}} &
   \rotatebox[origin=c]{90}{\textit{Counterfactual}} &
   \rotatebox[origin=c]{90}{\textit{Input Influence}} &
   \rotatebox[origin=c]{90}{\textit{Sensitivity}} &
   \rotatebox[origin=c]{90}{\textit{Case-based }}&
   \rotatebox[origin=c]{90}{\textit{Demographic}} &
   \rotatebox[origin=c]{90}{\textit{Conversational}} &
  \rotatebox[origin=c]{90}{\textit{Non-conversational}} &
  \rotatebox[origin=c]{90}{\textit{Model Agnostic}} &
   \rotatebox[origin=c]{90}{\textit{Model Specific}} &
  \rotatebox[origin=c]{90}{\textit{Goal-Driven}} &
  \rotatebox[origin=c]{90}{\textit{Data-Driven}} &  \rotatebox[origin=c]{90}{\textit{Local}} &
  \rotatebox[origin=c]{90}{\textit{Global}} & & &\\ \hline 
 Kim et al. \cite{kim2018textual}& \cmark &  &  &  &  & \cmark &  &  & \cmark &  & \cmark &  &  \cmark & \cmark &  & UG + ED & B, C & P, C%DV, AD, PA
 \\ \hline

Chakraborti et al. \cite{chakraborti2017plan} & \cmark & & & \cmark & & & & &\cmark &  & \cmark &\cmark &  & & \cmark & UG & B \& C & PL
%DV, AI, IN 
\\ \hline

Raman et al. \cite{raman2013sorry} & \cmark & & & \cmark & & & & &\cmark &  & \cmark &\cmark &  &  & \cmark & UG & B \& C & PL
% DV, AI, IN
\\ \hline

Xu et al. \cite{xu2020explainable} & \cmark &  &  &  &  & \cmark &  &  & \cmark &  & \cmark &  &  \cmark &  &  & UG &  A \& C & P
%PA, AD, SA, RG 
\\ \hline

Kim \& Canny \cite{kim2017interpretable}& \cmark &  & \cmark  &   &  & \cmark &  &  & \cmark &  & \cmark &  & \cmark  & \cmark  &  & UG & B \& A & P
%DV, AD, PA  
\\ \hline

Cultrera et al. \cite{cultrera2020explaining}& \cmark &  &  &  \cmark &  &  &  &  & \cmark &  & \cmark &  & \cmark  & \cmark  &  & UG & B \& A & P
%DV, AD, PA 
\\ \hline

Schneider et al. \cite{schneider2021explain}& \cmark & &  & \cmark  &  &  &  &  & \cmark  &  &  &  & \cmark  &  \cmark &  & ED & A \& C & P
%DV, SA, AI 
\\ \hline

% Neerincx et al. \cite{neerincx2018using}& \cmark & \cmark &  & \cmark  &  &  &  & \cmark &  &  & \cmark & \cmark &   &   & \cmark & UG & B \& C & P, PL
% %DV, SA, AI  
% \\ \hline

Rahimpour et al. \cite{rahimpour2019context}&  &  &  & \cmark &  &  &  &  & \cmark &  & \cmark &  &  \cmark & \cmark &  & UG & B \& C & P
%AD, DV, AI, AD 
\\ \hline

Shen et al. \cite{shen2020explain} & \cmark & &  & &  & \cmark &  &  & \cmark &  & \cmark & & \cmark  & \cmark  &  & ED & B & P
%DV
\\ \hline

Ben-Younes et al. \cite{ben2020driving}& \cmark &  &  &  &  & \cmark &  &  & \cmark &  &  \cmark &  &  \cmark & \cmark &  & UG & A \& C & P
%PA, AD, SA, RG
\\ \hline

Nahata et al.~\cite{nahata2021}& \cmark &  & \cmark & \cmark &  &  &  & \cmark &  & \cmark & &  & \cmark & \cmark  &  & UG &  B & PL
%PA, AD, DV 
\\ \hline

Ha et al. \cite{ha2020effects}& \cmark &  &  & \cmark &  &  &  &  & \cmark &  & & \cmark &   &   &  & ED + PC &  A \& B & P
%PA, AD, DV 
\\ \hline

Koo et al. \cite{koo2015did}& \cmark &  &  & \cmark &  &  &  &  & \cmark &  & & \cmark &   &   &  & ED + PC &  A \& B & P
%PA, AD, DV 
\\ \hline

% Cruz et al. \cite{cruz2019memory}& \cmark & \cmark &  & \cmark  &  &  &  & \cmark &  &  & \cmark & \cmark &   &   & \cmark & UG &  B \& C & P
% %DV, SA, AI  
% \\ \hline

Bojarski et al. \cite{bojarski2018visualbackprop}& \cmark &  &  &   &  & \cmark &  &  & \cmark &  & \cmark &  &  \cmark &  \cmark &  & UG &  B \& C & P
\\ \hline

Mori et al. \cite{mori2019visual}&  \cmark &  &  & \cmark  &  &  &  &  & \cmark &  & \cmark &  &  \cmark &  \cmark &  & UG &  B \& C & P
\\ \hline

Liu et al. \cite{liu2018impact}&  \cmark &  &  & \cmark  &  &  &  &  & \cmark &  &  & \cmark  &   &  \cmark &  & ED &  A & P
\\ \hline

Omeiza et al. \cite{omeiza2021not}&  \cmark & \cmark & \cmark & \cmark  &  &  &  &  & \cmark &  &  & \cmark &   &   &  & ED & A & P
\\ \hline

Rizzo et al. \cite{rizzo2019reinforcement}& \cmark &  &  & \cmark  &  &  &  &  & \cmark &  & \cmark &  &  \cmark &  \cmark &  & UG &  B & P
\\ \hline

Liu et al. \cite{liu2020interpretable}&  \cmark &  &  & \cmark  &  &  &  &  & \cmark &  & \cmark &  &  \cmark &  \cmark &  & UG &  B \& C & P
\\ \hline

Omeiza et al.~\cite{omeiza2021towards}&  \cmark & \cmark & \cmark & \cmark  &  &  &  &  & \cmark &  &  & \cmark &   &   &  & ED & A & P
\\ \hline

\end{tabular}
\label{tab:papers}
\end{table*}

The overview provided in Table~\ref{tab:papers} indicates that some types of explanations (e.g., sensitivity, demographics,  contrastive, counterfactual, model-agnostic, and global explanations) are rare in the autonomous driving literature. This may be due to the nascent nature of the explainable  autonomous driving domain.

\section {Explainable Autonomous Driving Operations}
\label{sec:expops}
\begin{figure}
  \includegraphics[width=9cm,height=9cm,keepaspectratio]{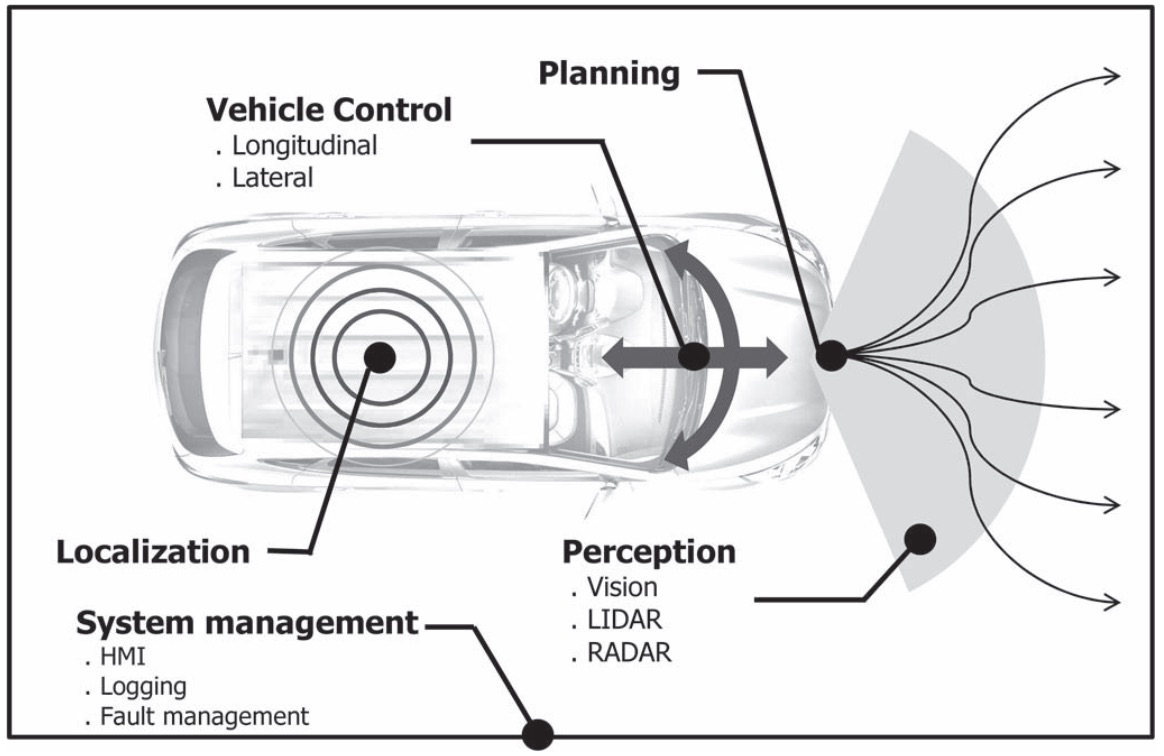}
  \caption{Key operations of an autonomous vehicle~\cite{jo2014development}. In Section~\ref{sec:expops} and Section~\ref{sec:hmi}, we discuss the role of explanations within these key operations.}
  \label{fig:av}
\end{figure}

This section provides a high-level description of the different operations of an AV and a review of previous work on explanation in each operation. The operations include perception, localisation, planning, control and navigation, and system management (which includes event data recorder and human-machine interaction) \cite{jo2014development}; see Figure \ref{fig:av}. 

\begin{figure}
  \includegraphics[width=\columnwidth,keepaspectratio]{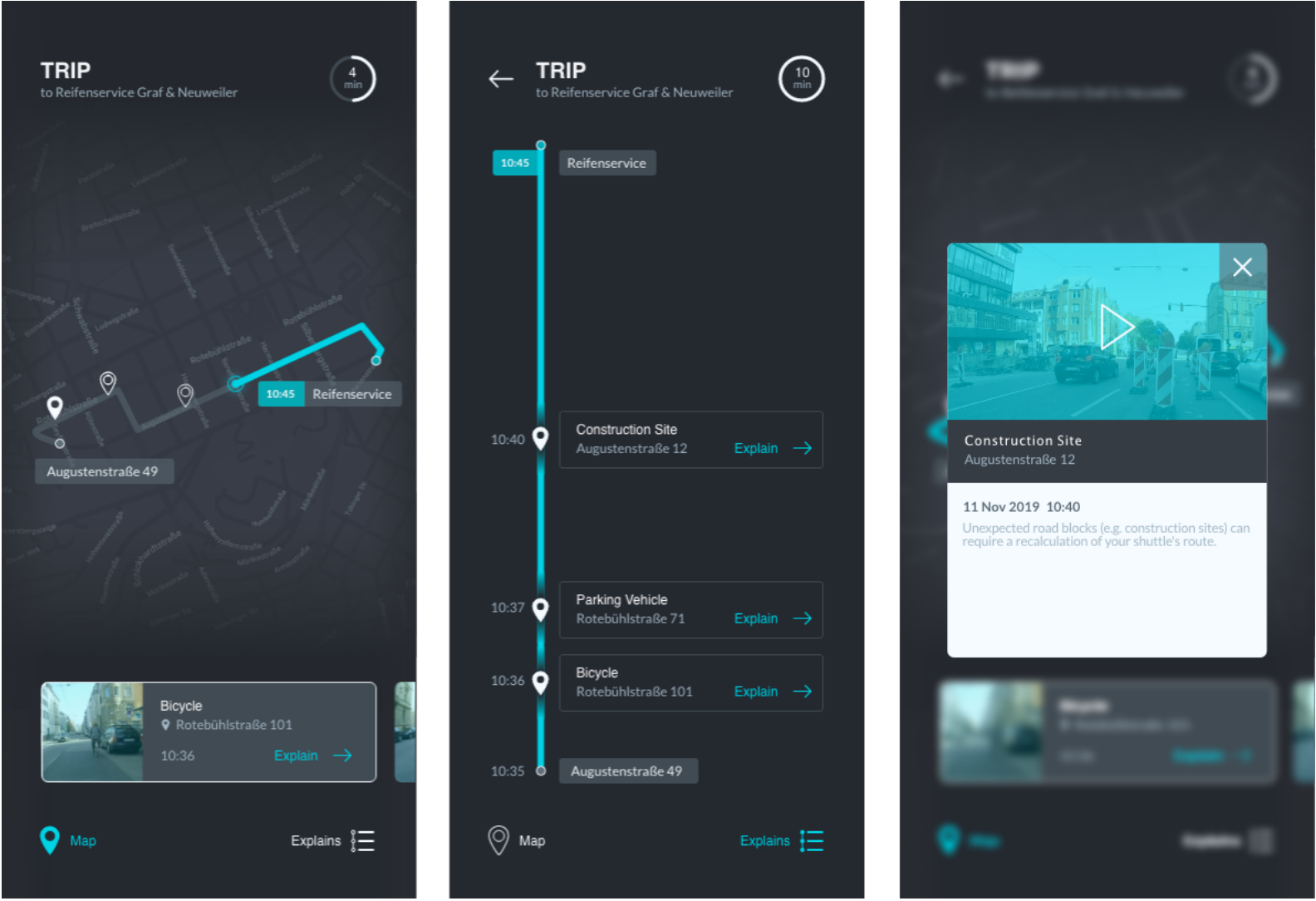}
 
  \caption{An example of a mobile interface for an AV explainer~\cite{schneider2021explain}. This interface is used for posthoc explanation provision. The app provides a record of a journey and can provide explanations in three different views at strategic points of the journey.}
  \label{fig:views}
\end{figure}

% Human-machine interaction is a crucial aspect of an AV could provide insights into explanation agents' interfaces. Thus, we omit human-machine interaction in this section and describe it in greater detail in Section \ref{sec:hmi}.

\subsection{Perception}
In this section, we identify various perception datasets that have been used and those that can potentially be used for explanation generation purposes. We also review recent data-driven explanation methods and previous works that have applied these explanation methods to the AV perception task.
\subsubsection {Driving Datasets For Posthoc Explanations}
Several driving datasets have been made available for the purpose of training machine learning models for autonomous vehicles (See~\cite{janai2020computer}). Some of these datasets have annotations---e.g., handcrafted explanations \cite{kim2018textual,you2020traffic}, vehicle trajectories \cite{houston2020one}, human driver behaviour \cite{ramanishka2018toward,shen2020explain} or anomaly identification with bounding boxes \cite{xu2020explainable,you2020traffic}---that are helpful for posthoc driving behaviour explanation. We have categorised the sensors used in collecting the datasets into exteroception and proprioception types, and the annotations in the datasets that are useful for developing explainable AVs. We also identified different stakeholders that can potentially benefit from the explanations. See Table~\ref{tab:datasets}). Although the datasets are helpful for developing explanation methods, it is important to note the potential challenges associated with the use of these datasets. Each dataset was collected from one region of the world, thus, chances are high that they may not generalise, especially where traffic signs, rules, and road topology are quite different to that of other regions; this can potentially lead to biased driving decisions. Also, most of the datasets only provide a video of the external environment and do not provide internal AV state data. It is therefore a concern as to whether the explanation techniques designed with this dataset will be very faithful to the AV. We further discuss these challenges in Section~\ref{sec:datasets}.
\subsubsection{Vision-Based Explanations for AVs}
\begin{figure}
  \includegraphics[width=\columnwidth,keepaspectratio]{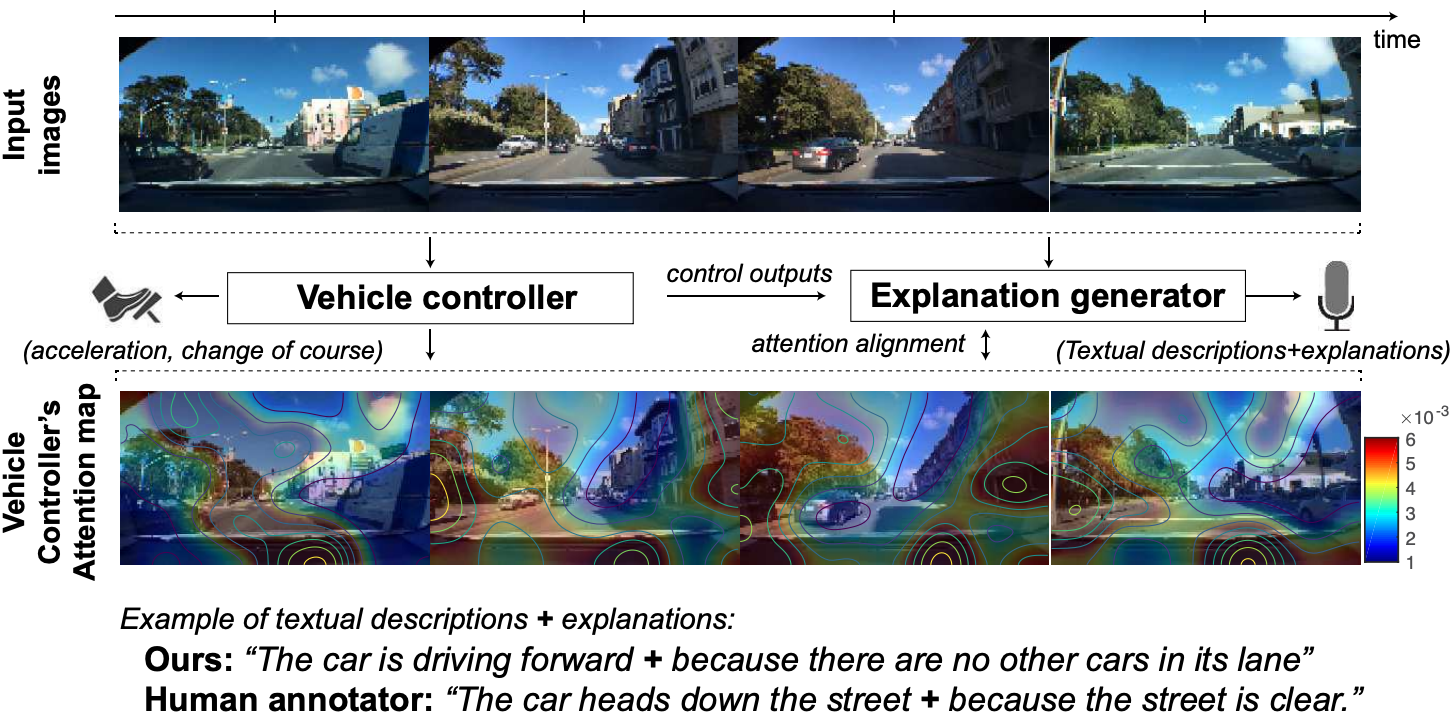}
 
  \caption{The vehicle control model predicts commands such as acceleration and a  change of course,  the explanation generator model generates natural language explanations and attention maps~\cite{kim2018textual}.}
  \label{fig:kim}
\end{figure}

Various methods have been proposed to explain neural networks which are fundamental structures for perception and scene understanding in AVs. Some of the prominent methods are \textit{gradient-based}. Gradient-based or backpropagation methods are generally used for explaining convolutional neural network models. The main logic of these methods is dependent on gradients that are backpropagated from the output prediction layer of the CNN back to the input layer \cite{das2020opportunities}. They are often presented in form of heatmaps (see Figure \ref{fig:kim}). These methods mainly fall under the input influence explanation style in the explanation categorisation presented in Table~\ref{tab:papers}.

% \cite{mundhenk2019efficient}. 
We provide some examples of gradient-based methods that are useful for explanations in AV perception. Refer to \cite{tjoa2019survey,zablocki2021explainability} for a survey on vision-based explanation methods.
\begin{itemize}
    \item Class Activation Map (CAM) \cite{zhou2016learning} and its variants like Gradient Class Activation Map (Grad-CAM) \cite{selvaraju2017grad}, Guided Grad-CAM \cite{tang2019interpretable}, Grad-CAM++ \cite{chattopadhay2018grad}, Smooth Grad-CAM++ \cite{omeiza2019smooth}
    
    \item Other gradient-based methods include VisualBackProp \cite{bojarski2018visualbackprop}, Layer-wise Relevance Propagation (LRP) \cite{lapuschkin2019unmasking, samek2016interpreting}, DeepLift \cite{shrikumar2017learning},
    \cite{zeiler2014visualizing}, and Guided-Backpropagation \cite{springenberg2014striving}.

\end{itemize}

Many of the vision-based explanations for AVs stem from generic gradient-based methods explained above. For example, 
Bojarski et al.~\cite{bojarski2018visualbackprop} proposed VisualBackProp for visualising super-pixels of an input image that is most influential to the predictions made by a CNN model. In~\cite{bojarski2018visualbackprop}, VisualBackProp on an end-to-end learning system for autonomous driving (PilotNet~\cite{bojarski2016end}) was applied to check whether the explanation method is able to show the parts of a driving scene image that are necessary for the steering operation of the AV model.

Kim et al.~\cite{kim2018textual} proposed an approach for explanation generation in autonomous driving. The approach involves training a convolutional neural network end-to-end from images to the vehicle control commands (which are acceleration and change of course). Further, textual explanations of the model actions are produced through an attention-based video-to-text model trained on the BDD-X dataset. Explanations were provided in form of saliency maps and text (see Figure \ref{fig:kim}). A related work by Xu et al.~\cite{xu2020explainable} focused on scene understanding, highlighting salient objects in input that can potentially lead to a hazard. These objects are described as action inducing since their state can influence the vehicles' decisions. Apart from identifying objects, a sequence of short explanations was generated.

\subsection{Localisation}
Localisation in AVs is the process of determining the pose (e.g., location and orientation) of the AV relative to
a given information (e.g., map) of the environment. Precise and robust localisation is critical for AVs in complex environments and scenarios \cite{wang2017map}. For effective planning and decision making, the position and orientation information is required to be precise in all weather and traffic conditions. One of the goals of a precise and robust localisation is to ensure that the AV is aware whether it is within its lane~\cite{reid2019localization} for safety purposes.
Safety is often considered the most important design requirement and it is critical in the derivation of requirements for AVs~ \cite{reid2019localization}. Hence, communicating position over time and with justifications as explanations is crucial to expose increasing error rates in a timely manner before they cause an accident. For instance, the position errors can be transmitted continuously through a wireless channel to an operation centre from which the AV is managed. An interface that displays this information (e.g., a special dashboard or mobile application as shown in~\cite{schneider2021explain}) is provided and it is able to trigger an alarm for immediate action (e.g., safe parking) when the error margin is exceeded.
Although there seems to be less research related to explainable localisation, intelligible explanations remain key. They would allow for easy communication of the position of an AV, including the measurement's precision and error \cite{reid2019localization}, of an autonomous vehicle during the localisation process in the form of clear and intelligible explanations. Explanations from localisation will be handy for Class B stakeholders (i.e., system developers) for debugging AVs because it can facilitate positional error correction and provide other stakeholders a perception of reliability and safety for AVs. Potentially, it will inform the development process of more robust localisation procedures.
\begin{table*}[]
\begin{center}
\caption{Driving datasets that can be used to develop explanation methods for AVs and the stakeholders that would potentially benefit from such explanations.}
\begin{tabular}{llllll}
\hline

\textbf{Dataset} & \textbf{Size} & \textbf{Exteroception} & \textbf{Proprioception } &  \textbf{Annotation \& Explanation} & \textbf {Stakeholders}   \\
&  &  Sensors (e.g., Cameras) & Sensors (e.g., CAN) &   & (see Sec.~\ref{sec:stakeholders}) \\\hline \\
\textbf{BDD-X \cite{kim2018textual}}   & 7K $\times$ 40s & \cmark & \xmark  & \parbox{5cm}{Textual \textit{Why explanation} associated with videos segments with heatmaps} & Class A, B, and C \\ \\

\textbf{BDD-OIA \cite{xu2020explainable}}   & 23K $\times$
 5s  & \cmark & \xmark  &  \parbox{5cm}{Actions and \textit{Why explanation}} & Class A, B and C\\ \\
\textbf{DoTA \cite{yao2020and}} & \parbox{1.5cm}{4,677
videos (73,193$s$)} & \cmark & \xmark & \parbox{5cm}{\textit{What explanation} (Temporal and spatial anomaly identification with bounding boxes)} & Class B and C \\\\ 
\textbf{CTA \cite{you2020traffic}} & 1,935 $\times$ 17.7$s$
 & \cmark & \xmark & \parbox{5cm}{Why explanation for accidents with cause and effects} & Class B and C\\ \\
\textbf{HDD \cite{ramanishka2018toward}} & 374,400$s$ & \cmark & \cmark & \parbox{3cm}{\textit{What explanations} for driver actions} & Class B\\ \\
\textbf{BDD-A Extended \cite{shen2020explain}} & $1,103 \times 
10s$ & \cmark & \xmark & \parbox{5cm}{Human gaze inciting \textit{why and/or what explanation}, explanation necessity score} & Class B \\\\

\textbf{Lyft Level5 \cite{houston2020one}} & 360,000$s$ & \cmark & \xmark & \parbox{5cm}{Trajectory annotation} & Class B \\\\ \hline
\end{tabular}
\label{tab:datasets}
\end{center}
\end{table*}
\subsection{Planning}
%cite Lars
Through AI planning and scheduling, the sequence of actions required for an agent to complete a task are generated. These action sequences are further utilised in influencing the agent's online decisions or behaviours with respect to the dynamics of the environment it operates in~\cite{ingrand2017deliberation}. The planning system is an important aspect of autonomous vehicles because of the complex maneuvers they make in dynamic, complex, and sometimes less structured or cluttered environments (e.g., urban roads, street roads with lots of pedestrians and other road participants). In fact, traffic elements (e.g., road side infrastructures, road network, road signs, and road quality) are dynamic and can change with time; this makes AVs regularly update their plans (and even learn sometimes) as they operate. 
Often, the amount of data (e.g., descriptions of objects, states, and locations) that the AV processes per time is larger than such that a human may be able to process, and continuously and accurately keep track of. Hence, a stakeholder riding in an AV may be left in a confused state when the AV updates its trajectory without providing an explanation.

Explainable planning can play a vital role in supporting users and improving their experiences when they interact with autonomous systems in complex decision-making procedures \cite{chakrabortiemerging}. According to \cite{sado2020explainable}, depending on the stakeholder, the process may involve the translation of the agent's plans into easily understandable forms, and the design of the user interfaces that facilitate this understanding. Relevant work include XAI-PLAN \cite{borgo2018towards}, WHY-PLAN \cite{korpan2018toward}, refinement-based planning (RBP) \cite{bidot2010verbal}, plan explicability and predictability \cite{zhang2017plan}, and plan explanation for model reconciliation \cite{chakraborti2017plan,chakraborti2019plan}. 

\paragraph{XAI-PLAN} is a domain-independent, planning system agnostic, and an explainable plan model that provides initial explanations for the decisions made by an agent planner \cite{borgo2018towards}. The user explores alternative actions in a plan and a comparison is done with the user's resulting plan and the plan that was suggested by the planner. The XAI-PLAN framework then provides an explanation to justify discrepancies. This kind of interaction encourages and enhances mixed-initiative planning which has the potential of improving the final plan. Interestingly, users can pose contrastive forms of queries in the form ''why does the
plan contain action X rather than action Y?”.
\paragraph{Refinement-based Planning (RBP)}
A related transparent and domain-independent framework called refinement-based planning (RBP) \cite{bidot2010verbal} produces explanations of verbal plans upon a verbal query from a user. It possesses an enhanced representation of the search space, providing a 2-way search (i.e., forward and backwards) capability when generating plans. This allows for flaw detection and plan update or optimization. Using states and action primitives, the RBP paradigm integrates partial-order causal-link planning and hierarchical planning \cite{biundo2014abstract} (hybrid planning
framework).

\paragraph{Why-Plan}
Korpan and Epstein \cite{korpan2018toward} also proposed Why-Plan, an explanation technique in human-machine collaborative planning. The method juxtaposes a person's and autonomous agent's objectives in a path planning process and provides explanations to justify the differences in panning objectives in a meaningful and human-friendly fashion. It basically addresses questions like "why does your plan involve that action?"

The explainable planning frameworks described above and the related work by \cite{chakraborti2017plan,chakraborti2019plan,hayes2017improving,neerincx2018using} can serve as basics to build upon for plan explanations in AVs.

\subsection{Vehicle Control}
Control in an AV generally has to do with the manipulations of vehicle motions such as lane changing, lane-keeping, and car following. These manipulations are broadly categorised under longitudinal control (speed regulation with throttle and brake) and lateral control (i.e., automatic steering to follow track reference) \cite{khodayari2010historical}. 

ADAS currently works based on the AV's sensor information obtained from observing the environment. Interfaces that come with ADAS now display rich digital maps~\cite{tomtom}, vehicle's position, and track related attributes ahead or around the vehicle. Stakeholders may issue investigatory queries when the AV makes a decision against their expectations. For instance, the stakeholder may want to ask different questions based on current contexts (e.g., near-miss, special vehicle case, or collision). Investigatory queries could be in form of a `why' question (e.g., `Why did you turn left?'), `why not' or contrastive question (e.g., `why did you switch to the left lane instead of the right lane'), `what if' or counterfactual questions (e.g., `what if you turned left instead of right?'), or `what' question (e.g., `What are you doing?'). 

Other than existing in-vehicle visual interfaces, such as mixed reality (MR) visualization \cite{sasai2015mr}, and other flexible (i.e., highly reconfigurable) dashboard panels~\cite{marques2011flexible}, in-vehicle interfaces that support the exchange of messages between the stakeholder and the AV is crucial. The user should be able to query the interface and receive explanations for navigation and control decisions in an appropriate form; either through voice, text, visual, gesture or a combination of any of these options.

In the next section, we address questions regarding explanation in relation to AV management and interaction with the respective stakeholders.

\section{System Management}
\label{sec:hmi}
In this section, we review works relating to event data recording (EDR) in AVs and human-machine interactions involving in-vehicle interfaces and external human-machine interfaces (eHMI) that could be potentially used for explanations.
\subsection{Logging and Fault Management: Event Data Recorder}

The event data recorder (EDR) serves as a recording device in automobiles to log information related to vehicle accidents. Upon a posthoc analysis, a better understanding of how certain faults or accidents come about is achieved \cite{wu2013driving}.

The installation of EDR in passenger vehicles has been a mandatory process in the United States since 2014. Recently, the National Transportation Safety Board (NTSB) suggested the need for risk mitigation pertaining to monitoring driver engagement and the need for better event data recording requirements for autonomous driving systems after the Tesla crash case in 2018~\cite{teslacrash}.

As autonomous vehicles increase in society and gain more public attention, it is necessary to discriminate human driver errors and negligence from the AV's errors (arising from non-adapted or poor product design or a product defect \cite{bose2014black, kohler2014current}) and express these errors in explanations. Martinesco et al.~\cite{martinesco2019note} attributed the existing challenge in ascribing faults to the appropriate traffic participant to the difficulty in identifying and evaluating the correct cause of an accident.

In line with this, the National Highway Traffic Safety Administration (NHTSA) calls for the industry and standard bodies such as SAE and IEEE to develop a uniform approach to address data recording and sharing\footnote{See relevant documents here: https://www.nhtsa.gov/fmvss/event-data-recorders-edrs} which may, in turn, be useful for explanations. Pinter et al.~\cite{pinter2020road} deplored the inability of the existing EDRs to provide sufficient data needed to reconstruct the behaviour of a vehicle before and after an accident, and to a degree that the accident could be analysed from the perspective of liability. As AV functions continue to increase (eventually leading to full autonomy), the storage of a satisfactory number of parameters is needed for the reconstruction of the vehicle's behaviour  and the provision of explanations for a reasonable amount of time before and after the accident becomes crucial.

As an effort towards building more effective EDRs that can support explanation provision, different approaches, which include the use of blockchain technologies, and more effective and robust data models have been proposed. Guo et al.~\cite{guo2018blockchain} proposed a blockchain-inspired EDR system for autonomous vehicles to achieve indisputable accident forensics by providing trustability and verifiability assurance of an event's information. With this blockchain approach, the verification and confirmation of a new block of event data are possible with no central authority involved.
In terms of storage mechanisms and reliability, Yao et al.~\cite{yao2020smart} proposed a Smart Black Box (SBB) to
supplement traditional data recording with value-driven higher-bandwidth data capture. The SBB uses a deterministic mealy machine~\cite{HARRIS2013108} based on data value and similarity to cache short-term histories of data as buffers. By optimising value and storage cost trade-offs, the appropriate compression quality for each data chunk in the driving history data is determined. Prioritised data recording prevents the retention of low-value buffers. By discarding them, space is made available to store new data.

With the EU upcoming legislative rules on EDR beginning 2022~\cite{eclaw} (and a similar one in China \cite{chinalaw}), there is the question as to whether existing data storage facilities are sufficient for the data needs for accident investigations involving automated vehicles. For efficient storage space management, a well defined data package which puts the data points (with necessary parameters) and the frequency of measuring and recording that can enable full reconstruction of AVs' regular and irregular movements is necessary for event explanation purposes. The data model from Pinter et al.~\cite{pinter2020road} can be used to determine the data content required in an EDR, sufficient for accident investigations, and suitable for vehicles at different autonomy levels. Further, Bohm et al.~\cite{bohm2020new} proposed a broader database in relation to the US EDR regulation (NHTSA 49 CFR Part 563.7) after carrying out a study involving the reconstruction of real accidents with ADAS enabled vehicles to investigate requirements. These advancements in EDRs are relevant for the development of explanation techniques for accidents (and other critical events). It may also draw researchers' attention to explainable EDR which is currently very much under-explored. Human-machine interaction (HMI) is a key aspect of explanation in AVs. In the next section, we will discuss the relationship between HMI and explanations in the autonomous driving context.

\subsection{Human-Machine Interaction}
AVs possess components for sensing, decision making, and the operation of the vehicles, requiring minimal human driving~\cite{smith2015automated}. They can operate in complex environments where the decision set is large~\cite{yurtsever2020survey}. This poses a challenge to the understandability of their operational modes. AVs are seen to have evolved over the years in terms of automation level, and in-vehicle technologies and interfaces (i.e., technologies and interfaces within the vehicle). Essentially, vehicles in the SAE levels 0 to 2 have a low explanation requirement due to their low complexity. For vehicles in levels 3 and above, the explanation requirement is high due to their high complexity. Table~\ref{tab:sae} provides a summary based on~\cite{ddash1, ddash2}, with SAE levels and explanation requirements. As shown in Table~\ref{tab:sae}, vehicle instrument (i.e., an instrument that measures some quantities about the vehicle) interfaces evolved to adaptive displays where content is presented in a form that enhances user experience, and with enhanced positioning features, e.g., global navigation satellite system (GNSS).

\begin{table*}[]
 \caption{Vehicle Instrument Interface Evolution and Explanation Need}
\begin{tabular}{lp{5cm}llp{6.6cm}}
\toprule
\# & Explanation Interface     & SAE Automation Level & XAI Demand & Vehicle Examples \\
\midrule
1 & Fully analog interface & Level 0      & Low          & Old Ford vehicles and similar vehicles back before the year 1990   \\ 
2 & Partly analog and digital interface (e.g., digital odometer, analog speed dial) & Level 0       & Low        & Older Honda Civics, Citroen C4 Picasso and others mostly between 1990 and 2000.
      \\ 
3 & Mostly digital interface & Level 0 and 1      & Low          & BMW 5 Series, Fiat 500, and Jaguar XF and others mostly between 2010 and 2016
  \\ 
4 & Fully digital interface with adaptive display, GNSS) & Level 2 and 3 & Moderate               & Tesla Autopilot, Audi A8 2016 to present \\ 
5 & Fully digital interface with adaptive display, Sat Nav) & Level 4 & High & Waymo cars 2016 to present  \\
\addlinespace
\bottomrule
\end{tabular}
\label{tab:sae}
\end{table*}

Recent highly automated vehicles are incorporating more enhanced interaction technologies. Moreover, novel interaction technologies provide the opportunity for the design of useful and attractive in-vehicle user interfaces that abstract and explain vehicle automation operations (e.g., perception, planning, localisation, and control) exist. In the next section, we will discuss some research on in-vehicle interfaces.

\subsubsection{Novel Interaction Technologies}
The in-vehicle user interface is essential for efficient explanation provision, and hence, enhancing the driving experience~\cite{schmidt2010automotive}. There are studies that suggest that interface design trends impact driving experience. For example, Jung et~al.~\cite{jung2015displayed} explored the impact of the displayed precision of instrumentation estimates of range and battery state-of-charge on drivers' driving experience, and attitude towards varying conditions of resource availability in an all-electric vehicle. Results from the study showed that it can be advantageous to display the uncertainty values associated with a measure rather than concealing it as participants presented with an ambiguous display of range measure reported a preserved trust level towards the vehicle. Although presenting users with a single number value increased reading and apprehension time, the implication of disguised uncertainty on user experience and behaviour has to be carefully considered in critical situations. 

A related work by Mashko et al.~\cite{mashko2016virtual} involved the assessment of in-vehicle navigation systems with a visual display where virtual traffic signs were represented on an in-vehicle display to assist better orient at road sections loaded with excess information clutter. The use of virtual traffic signs in-vehicle improved the drivers' concentration and reaction to traffic signs on the road. Langois~\cite{langlois2013adas} proposed an interface (Lighting Peripheral Display––LPD) that creates signals that are able to be handled by peripheral vision (the ability to see objects and movement outside of the direct line of vision) while driving in order to enhance the utility of ADAS. The LPD possessed a box illuminated by light-emitting diodes (LEDs) and reflected onto the windscreen. User tests conducted showed that driving performance and comfort were enhanced by LPDs. Sirkin et al.~\cite{sirkin2017toward} developed Daze, a technique for measuring situation awareness through real-time, in-situ event alerts. The technique is ecologically valid in that it is very similar in look and feel to the applications used by people in the actual driving environments, and can be applied in simulators and also in on-road research settings. The authors conducted a study that included simulated based and on-road test deployments in order to provide assurance that Daze could characterise drivers' awareness of their immediate environment and also understand the practicality of its use.

Having examined the existing interaction technologies, it would be worthwhile to look at what users actually prefer.
\subsubsection{In-vehicle Interfaces---User Preferences}
Learning about the experiences of in-vehicle participants will help to inform what users' preferences are.
Mok et al.~\cite{mok2015understanding} described a Wizard of Oz study to get insights into how automated vehicles ought to interact with human drivers. Design improvisation sessions were conducted inside a driving simulator with interaction and interface design experts. While the two human operators (wizards) controlled the audio and driving behaviour of the car, the participants were driven through a simulated track with different terrain and road conditions. The study noted that: 
\begin{enumerate}
    \item instead of taking over full control, participants wanted to share control with the vehicle;
    \item participants like to know exactly when a handover (mode switch) happens and require a clear alert from the vehicle to that effect;
    \item to the participants, delayed responses and unperformed requests were acceptable as long as the responses provided are correct/proper;
    \item AVs have a variety of means to help sustain or improve participants' trust in them.
\end{enumerate}

Fu et al.~\cite{fu2020too} studied the effect of varying sensitivity and automation levels of in-vehicle collision avoidance systems. The authors explored this with the automatic emergency braking (AEB) systems in level 3 autonomous vehicles, for which the attention of the driver is needed to monitor the system for failures. Drivers reacted more (in terms of vigilance and awareness) to the system when it was biased to under-report hazards. The result also suggested that higher levels of autonomy in vehicles result in a lower level of driver vigilance and awareness. This was discovered when the roles of the driver and the computer were reversed, where the driver was meant to supervise an imperfect higher-level automated system; the driver's reaction performance worsened during a critical event. Related studies are described in \cite{fu2020too, mok2015emergency}.

Park et al.~\cite{park2020driver} conducted a study in an attempt to understand the extent to which semi-AV decision-making should account for individual user preference. Having considered 18 different scenarios with tactical driving goals, significant differences were discovered in scenario interpretations, AV perceptions, and vehicle decision references. The alignment of individual preference with AV decision yielded more positive changes in the impression of the vehicle than unaligned decisions.

Closely related to explanations, Ha et al.~\cite{ha2020effects}, Koo~et~ al.~\cite{koo2015did}, Omeiza et al.~\cite{omeiza2021not} investigated the effect of explanations on trust through empirical user studies. Ha et al.~\cite{ha2020effects} examined two explanation types, simple and attributional, as well as perceived risk on trust in AVs in four autonomous driving scenarios with varying levels of risk using a simulation of an in-vehicle experience. Their results indicated that an explanation type can greatly affect trust in autonomous vehicles and that under high levels of perceived risk, attributional explanations lead to the highest trust in AVs. 

Further, Schneider~et al~\cite{schneider2021explain} investigated whether the provision of explanations in simulated driving can enhance the user experience and increase the subjective feeling of safety and control. The provision of explanations did not influence user experience during and after the ride. See an example of a basic mobile explanation interface in~Figure~\ref{fig:views}. 

These works highlight the importance of the user-centred approach (which could have a great influence on the future of human-machine interaction) in the design and development of explanation methods for AVs.
In-vehicle user interfaces are a medium for the provision of visual, text, or voice explanations. However, previous works only focus on providing information about the vehicle to users without communicating reasons and/or causal links for decisions. This remains an open challenge for AV designers and researchers.
We now explore the interaction between an AV and other external traffic participants.

\subsubsection{AV and External Agents Interaction}
There are different categories of traffic participants that an AV has to interact with, and there are many studies focusing on the interactions between AVs and other traffic participants~\cite{eby2016use,yang2014vehicle}. Traffic participants that AVs will frequently interact with are: pedestrians, cyclists and other vehicles. 

\paragraph{Pedestrians}
Pedestrians and human drivers communicate intents to each other to inform the next choice of action~\cite{vsucha2014road,wilde1980immediate,sun2003modeling}. Autonomous vehicles should also have modalities for communicating intents to pedestrians~\cite{rasouli2019autonomous}.
Mahadevan et~al.~\cite{mahadevan2018communicating} conducted a study to get insights into interface designs that explicitly communicate autonomous vehicle awareness and intent to pedestrians. Different interface prototypes were developed and deployed in a study that involved a Segway and a car in a simulation setting. Results suggest that interfaces communicating vehicle awareness and intent can assist pedestrians attempting to cross at crosswalks and can exist in the environment outside of the vehicle. They suggested a combination of modalities (e.g., visual, auditory, and physical) in the interfaces.

Faas et al.~\cite{m2021calibrating} investigated pedestrians’ trust and crossing behaviour in repeated encounters with AVs in a video-based laboratory study. The occurrence of AV malfunction and system transparency with status and intent eHMI were studied. Their results showed that trust increases with the presence of status and intent eHMI and decreases when there is a malfunction in the AV but recovers quickly. Crossing onset time also decreased with the provision of the eHMI.  Crossing onset time indicates the time in seconds between the vehicle yielding and the
pedestrian stepping off the sidewalk~\cite{faas2020efficient}. It was noted that status eHMI can cause pedestrians to overtrust AVs, therefore, intent messages are needed to complement status eHMIs.

Pedestrians reactions to a ghost driver have also been investigated in the academic literature. A ghost driver in this context refers to a driver who pretends to be absent in the car even when they are in control of the driving operations. Moore et al.~\cite{moore2019case} conducted a Wizard-of-Oz driverless vehicle study aimed to test pedestrians' reactions to everyday traffic in the absence of an explicit eHMI. Although some pedestrians were surprised by the vehicle's supposed autonomy, others neither noticed nor paid attention to its autonomous nature. All the pedestrians crossed in front of the vehicle without explicit signalling. This suggests that the vehicle's implicit eHMI (which is basically its observed motion) may suffice. Therefore, pedestrians may not need the explicit eHMI in their interaction routine. A similar study by Moore~et~al.~\cite{moore2019visualizing} indicated that pedestrians crossed in front of a ghost vehicle with little hesitation even when the vehicle did not give any signal beyond its motion. However, Li~et~al.~\cite{li2020road} findings contradicts this claim by confirming pedestrians behaviours are different on encountering a vehicle with a hidden driver based on a study carried out in Europe.
% \end{itemize}

\paragraph{Interaction with Pedestrians with Reduced Mobility (PRM)}
Pedestrians with reduced mobility might need their support devices re-engineered to allow for effective interaction with AVs~\cite{asha2020views}. Pasha et al.~\cite{asha2020views} carried out a design study to explore interface designs for interaction between AVs and PRMs. The results from the analysis disclosed that visual cues are the most important interface elements, and street infrastructures are the most important location for housing cues for this category of pedestrians. They also found that wheelchairs might require an interface, and the current wheelchairs would have to be altered to allow for this interface.

\paragraph{Other Road-Participants}
Vehicle-cyclist interaction is an important topic to examine, especially in an environment where cycling is common. Cyclists and drivers currently communicate through implicit cues (vehicle motion) and explicit but imprecise signals such as horns, lights, and hand gestures \cite{hou2020autonomous}. Hou et al.~\cite{hou2020autonomous} designed a virtual reality (VR) AV-cyclist immersive simulator and a number of AV-cyclist interfaces to explore interactions between AVs and cyclists. Findings suggest that AV-cyclist interfaces can improve rider confidence in lane merging scenarios. Future AVs could consistently communicate feedback (explanations) which includes awareness and intent based on their sensor data~\cite{hou2020autonomous}.

In general, more research is needed to explore how the findings from these studies can be utilised to create effective and efficient interaction interfaces between AVs and stakeholders in order to facilitate the provision of explanations. In the next section, we will present some challenges around explainability in autonomous driving and suggest future research directions.
\begin{figure*}
  \includegraphics[width=\textwidth,keepaspectratio]{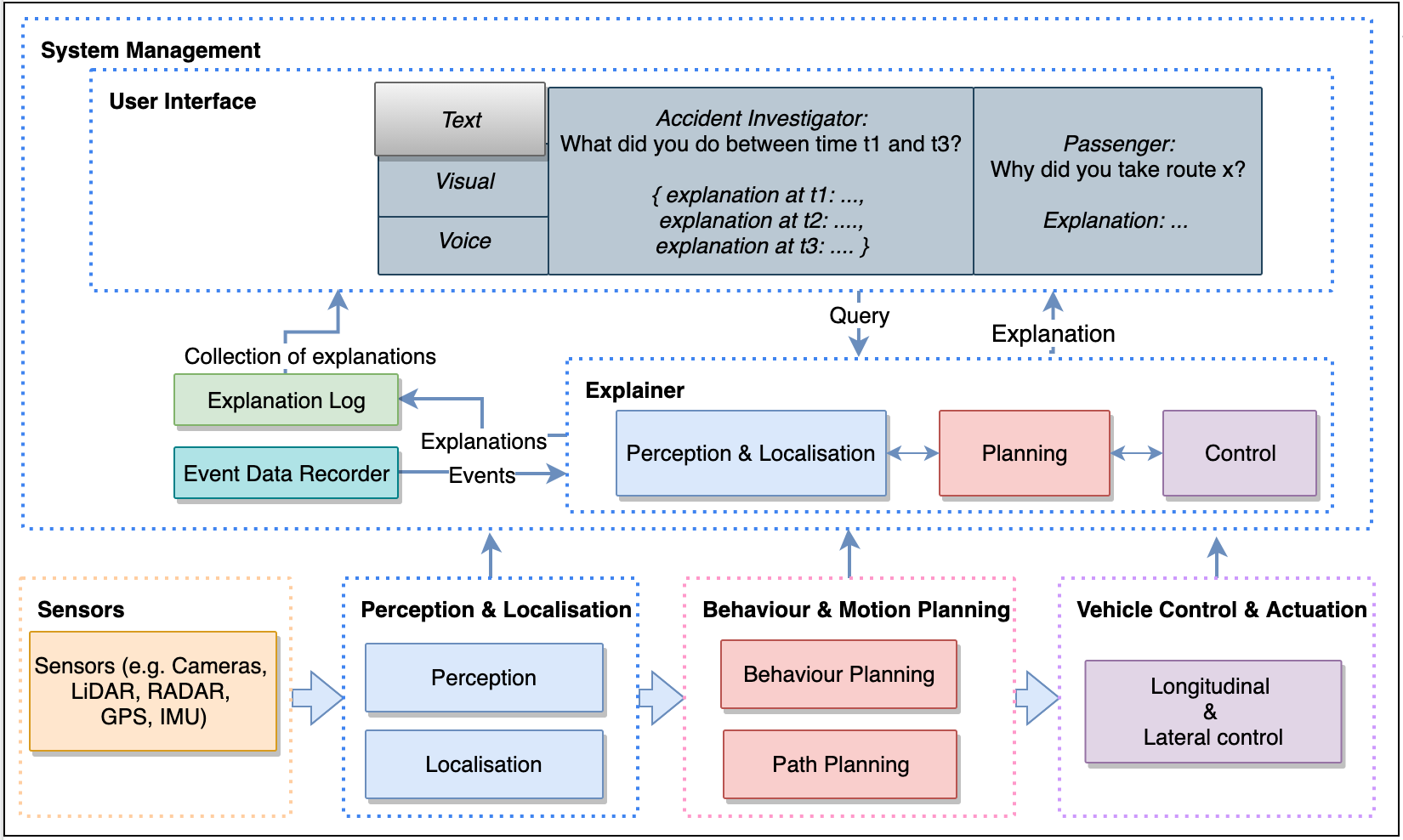}
 
 \caption{A conceptual framework for explainable AV following the perception, planning, and control paradigm. The explainer receives data from the perception and localisation system, planning system, and vehicle control system. This data can either be stored with time stamps and provided all at once as an explanation log to accident investigators, or each explanation is provided as it is generated. The framework provides a multi-modal interface for interacting with the explainer.}
  \label{fig:explainer}
\end{figure*}
\section{Challenges and Future Research Directions}
In this section, we discuss the challenges related to standards and regulations, user-centric explanation design and evaluation, and ethics. We also suggest means to address some of the key challenges.
\label{sec:challenges}
\subsection{Standards and Regulations}
Some of the standards provided in Table~\ref{tab:stds} are very relevant to explainability in autonomous driving. For example, ISO/TR 21707:2008 which specifies a set of standard terminology for defining the quality of data being exchanged between data suppliers and data consumers in the ITS domain is very relevant to AV explainability although not originally intended for explainability. While the quality of data is important, the presentation style and language and the interfaces by which the data is provided are also critical for explanations in autonomous driving. We suggest that this standard and others in Table~\ref{tab:stds} be explored for the development of more AV explainability related ones, and should be made easily accessible. 

Regulations regarding the explainability of automated systems are being set by countries and regions. However, these regulations seem to be too general and do not directly specify requirements for specific technologies and stakeholders, especially in autonomous driving. For example, the UK's ethics, transparency and accountability framework for automated decision-making~\cite{ukethics} states that: 
\begin{quote}
    When automated or algorithmic systems assist a decision made by an accountable officer, you should be able to explain how the system reached that decision or suggested decision in plain English. The explanation needs to be appropriate for your audience, expert or non-expert, and should be scrutinised and iterated by a multidisciplinary and diverse team (including end-users) to avoid bias and group speak.
\end{quote}

In a preliminary consultation paper on autonomous vehicles~\cite{npl}, the UK law commission states the recommendation of the National Physical Laboratory on explainability in autonomous driving:

\begin{quote}
It is recommended that the autonomous decision-making systems should be available, and able, to be interrogated post-incident. Similar to GDPR, decisions by automated systems must be explainable and key data streams stored in the run-up, during and after an accident.
\end{quote}

Explanations can help in assessing and rationalising the actions of an AV (\textit{outcome-based}), and in providing information on the governance of the AV across its design, deployment, and management (\textit{processed-based}). This is in line with the information commission office (ICO) guidelines~\cite{ico} for general AI systems. We suggest that regulatory guidelines for AV explainability should be set in line with these two goals. We have carefully adapted the ICO guidelines and the ITU’s comments on the `Consultation Paper 3
A regulatory framework for automated vehicles' to provide suggestions that include a conceptual framework for better explainability in AVs.

\subsubsection{Outcome-based explainability}
To be able to rationalise/explain the outcomes of an AV, explainability at the perception level (i.e., what the AV 'sees'), decision level (i.e., how the AV plans paths and motion) and action or control level (i.e., how the AV acts on its plan or how it executes its decisions) must be considered. We use these high-level terminologies: perception, decision, and actions to describe how explainability regulations could be made specifically for AVs.
\paragraph{Perception} Explanations for a perception system can come in two forms:
\begin{enumerate}
    \item An explanation capable of explaining the algorithms or software processes used to transform sensor data into a digital representation of the real-world and justification for such algorithms. In line with the Molly problem earlier mentioned in~\ref{sec:sec2}, it should be possible to obtain information on how Molly was represented digitally (sensor types and data transformations), and information about the circumstance, e.g., her location, position (maybe coordinates), and the time the representation occurred.
    \item An explanation that provides information on what this digital representation contains (e.g., pedestrian, vehicles, road fabric) and the state of these objects (e.g., crossing, heading north, static). For example, if Molly was detected, it should be possible to receive information to explain and justify the process (e.g., detection and tracking algorithm or software) by which Molly was detected and tracked and the detection and tracking confidence levels.
\end{enumerate}

In explaining outputs from a perception system, it is suggested that AVs should be able to provide real-time data access (both onboard or remotely) to their digital representation of the 3D world (including semantic information) when requested by authorised entities.

\paragraph{Decision} This involves the provision of insights into the planning operations (behaviour planning and path planning) of the AV. The decision-making steps involve planning paths and motion/behaviour based on observations (through perception) from the environment and its structured knowledge about the environment. These planning require reasoning and decision-making under several constraints, e.g, uncertainty about both the current and future state of the environment. It also involves the identification of potential risks, evaluating them and finding measures to mitigate the risks.

These uncertainties are associated with the confidence level in the AV’s predictions. AVs use prediction algorithms to predict their trajectory and that of other road participants, and the risk of collision associated with different plans.
In relation to the Molly Problem, to ascertain whether the AV had a good awareness of its environment upon which it made a decision, we must consider whether it was aware of the confidence levels of the models that detected and tracked Molly. It should be able to provide information on how confident it was about the subsequent steps or actions of Molly e.g., the probability that Molly will increase her speed in the next seconds. It should also be able to provide information about the considered plans (including the chosen ones), and the risk values associated with the considered plans.

It is suggested that when requested by authorised entities, AVs should be able to provide real-time data access (both in-vehicle or remotely) to the levels of uncertainty associated with its current and possible future digital representations of its environment and the uncertainty threshold upon which a plan or a risk mitigation action was selected.

\paragraph{Action} 
This deals with the provision of information to provide insight into the execution of the AV's plans for given contexts. It is the resulting vehicle dynamics to continuous control inputs in response to circumstances and situations observed. They are measurable outputs of the perception and decision-making steps that provide valuable insights into driving behaviour and risk. As such, the continual monitoring of actions can be used for assessing the behaviour of the AV in given circumstances.
We suggest that AVs should be able to provide real-time data access (both in-vehicle or remotely from the vehicle) to the actions of the AV, with respect to observations and knowledge; and the resultant decisions made, when requested by authorised entities.

\subsubsection{Process-based Explainability}

Process-based explainability in an AV is concerned with the provision of information that facilitates the independent assessment of the entire operations and governance of the AV. Process-based explainability takes perception, decision, and action data, including the governance processes of the entire AV operation. This makes it possible to reconstruct an event or accident immediately after it happens, significantly reducing the time to provide recommendations for future improvements.

Process explainability can be useful for fairness, safety and performance assessment, accountability and responsibility, and impact assessment. For fairness, the explanation outlines the steps taken across the design and the implementation of the AV to ensure that its decisions are generally unbiased and fair and whether someone has been treated equitably. For example, whether adequate measures were implemented to provide PRMs more time to cross the road; whether the pedestrian detection/classification system works with the same accuracy for people of colour; or whether predicted crash outcomes are representative of all members of the population, e.g, all body types, not just typical adult males.

For safety and performance evaluation, the explanation provides information on the steps taken across the design and implementation of the AV to maximise the accuracy, reliability, security, and robustness of its decisions and behaviours. For accountability and responsibility, the explanation provides insight on who is involved in the development and management of an AV, and who to contact for a human review of a decision. For impact assessment, the explanation provides information on the steps taken across the design and implementation of AV technologies. It considers and monitors the impacts that the use of the AV and its decisions has or may have on an individual, and on wider society.

It is suggested that real-time access to perception, decision, and action data and information about the process management (both in-vehicle or remotely from the vehicle) be made available for independent real-time processing to authorised authorities. Powers should also be granted to relevant authorities to impose sanctions in real-time based upon failure to meet explainability requirements.

In the next section, we suggest how the different AV operations fit into the AV explainability recommendations provided and we also provide a conceptual design to represent this.

\subsubsection{Conceptual Framework for Explainable AV}
We suggest a conceptual design of how we envision the core AV operations and components fitting together for explanation provisions purposes (see~Figure~\ref{fig:explainer}). The perception and localisation operations are fused together as both receive input directly from sensors (LiDAR, radar, GPS, IMU). The perception system provides a digital 3D representation of the world/environment including information about object detections, detection and tracking uncertainties, and object location, and passes them to the behaviour and motion planning system which makes decisions by estimating risks, finding mitigation actions, and producing a trajectory. The output from the behaviour and motion planning system informs the vehicle control and actuation system on the kind of inputs to send to the actuators to alter the vehicle dynamics to achieve some effect (e.g., reacting to an observation). Each of these operations feed directly to the explainer and the EDR in the system management compartment. The EDR organises its data and can provide its organised/structured data to the explainer. The explainer can generate explanations from a combination of data it has from the different operations and the EDR. Explanations can be requested in bulk in the form of an explanation log by specifying a time range or they can be requested in a conversational manner (i.e., one query at a time). The explanation log can provide process-based information that is useful for accident investigation and effective AV governance. The conversational style explanations are useful for outcome-based explanations as events occur.

\subsection{User-Centric Explanation Design and Evaluation}
%Explanation personalisation
\subsubsection{Trust}
Many of the existing works examining explanations in the context of autonomous driving do so using the trust objective. These works involved user studies with either synthetic laboratory tasks with little or no direct connection to the real system or used \textit{microworlds}~\cite{lee2004trust}. A microworld is a simplified version of a real system in which the critical elements are present but with the complexities eliminated to assist easy control of the experiment~\cite{brehmer1993experiments}. Attributes such as trust which have been used for explanation assessment in AVs \cite{koo2015did, ha2020effects} should be measured using a progressive experiment occurring in a real-world setting in order to obtain accurate results. This recommendation align with the trust node of the UKRI Trustworthy Autonomous Systems programme \footnote{https://www.ukri.org/news/new-trustworthy-autonomous-systems-projects-launched/} which stresses the need for  extensive human and autonomous systems interaction studies in relation to trust.

\subsubsection{Stakeholders' Consideration} As seen in Table~\ref{tab:papers}, research on explanations in AVs has mainly focused on the theory and implementation of explanations based on perception data with less user-centric empirical studies. Only a few user studies were conducted to elicit the requirements (e.g., when an explanation is needed and the appropriate explanations for each scenario) of the different stakeholders (e.g., some vulnerable members in Class A). Thus, the consideration of stakeholders in terms of explanation utility needs more attention. Further, most of the explanation methods proposed do not fulfil the properties of the standard causal explanations (e.g., that explanations should be contrasted with relevant foils) \cite{miller}. We suggest a more interdisciplinary approach to explanations in autonomous driving, where expert ideas from behavioural sciences (e.g., conversation theories), human-computer interaction (e.g., user interface designs), artificial intelligence, ethics (e.g., unveiling bias through explanations), philosophy (e.g., causal explanation theories) and psychology (e.g., folk psychology) are harnessed. Further, explanation techniques developed should be tested either through user studies, field trials, or laboratory studies to ensure that they fulfil requirements for the respective stakeholders they are designed for.
\subsection{Ethics: AV Operations}
\label{sec:datasets}
\subsubsection{Faithfulness and Blackboxness}
A major limitation in the existing work is the low assurance of faithfulness of the explanations generated. Intermediate data or states data of AV operations are currently missing in nearly all of the existing autonomous driving datasets. Also, the existing work on explainability in AVs are merely explanations provided by humans trying to rationalise the behaviour of an AV or an ego vehicle. These explanations obtained from rationalisation are used to annotate driving datasets which are subsequently used to train a deep explanation model. The complexity or blackbox nature of this process makes auditing difficult and thereby presents a challenge in the evaluation of faithfulness.

Considering faithfulness on a non-binary scale as suggested in \cite{jacovi2020towards}, explanations for autonomous driving behaviours could attain improved faithfulness when considering data from the different intermediate operations such as planning.
Constructing an AV’s navigation plans and exploring them is possible when the start and end goals of the AV are known beforehand. This information can be difficult to assess from AVs at present but, as a starting point, driving simulators and the Lyft Level5 dataset are helpful to create a prototype at the least. To ensure the intelligibility of explanations, the representation of these navigation plans need to be clear to lay users. Interpretable techniques such as navigation graphs and decision trees could be helpful in this regard. Navigation graphs and decision trees are inherently transparent, hence, meaningful interpretations and explanations (e.g., in natural language) can be easily obtained.

Also, for intelligibility, high-level commands (e.g., turning right, lane change left, and lead vehicle accelerating, among others) as used in the National Highway Traffic Safety Administration (NHTSA) report~\cite{najm2007pre} can be used to represent transitions between road and lane segments, and interaction with other road participants. 

The Sense Access eXplain (SAX) project\footnote{https://ori.ox.ac.uk/projects/sense-assess-explain-sax/} in the Oxford Robotics Institute (ORI) is currently collecting driving data which includes some proprioception information from the CAN bus of a Jaguar Land Rover (JLR) ego vehicle \cite{gadd2020sense}. Some of the relevant CAN bus data include wheel angle, yaw, acceleration, and braking, among others. The JLR is also equipped with in-vehicle microphones which record the spoken thoughts of the driving instructor who drives the vehicles and the passengers during field trials. Using this data to supplement exteroception data and navigation plans with interpretable representations and algorithms will help to enhance the faithfulness of explanations. Moreover, the recorded spoken thoughts of the JLR driver and passengers obtained during the trials can be further explored (e.g., through conversation analysis) to gain insights into how natural language explanations should be provided.

\subsubsection{Explanation Bias}
Another serious limitation in current research is the inherent bias existing in the resulting explanation models built on the perception datasets (presented in Table~\ref{tab:datasets}). The datasets are finite and are non-representative as they do not cover all possible driving scenarios in fair ratios. For instance, some traffic rules and signs differ with regions or continents. In addition to finite training data, it can be argued that the specification of the (finite-complexity) model (e.g., which variables to include and which to ignore) is still an open research question. How do we know that the AV's model of it environment at a particular time captures all of the variables that can influence its actions at that instance? Hence, the mis-specification of the model and the non-representative nature of driving datasets can introduce biases that propagate down into explanations. This challenge opens up new research questions around driving data quality, and the  adaptability and correctness of the AV's models.

There are hopes that AVs will bring many benefits to society. If these hopes must come through, the research community should take actions toward better research inclusivity (e.g., of researchers, regulators, and end-users). AV regulators can leverage on the suggestions outlined in this paper to develop more AV explainability related guidelines and regulations to address the highlighted issues.
\section{Conclusion}
\label{sec:conclude}
We have presented a literature survey on explanations for autonomous driving. We discussed the need for explanations in autonomous vehicles, and identified and categorised different autonomous driving stakeholders who would potentially utilise explanations. We identified and categorised relevant regulations and standards that could facilitate explanations. Having conducted a thorough review of the existing academic literature on explanations, we provided a categorisation for explanations and positioned the reviewed academic articles in the appropriate category. Consequently, we identified key challenges around explainability in autonomous driving. Some of the challenges include the absence of concrete regulations on AV explainability, the insufficient levels of interdisciplinary research around AV explainability, and bias and faithfulness issues potentially arising from the technologies built upon the current AV datasets. We provided recommendations for the development of AV-specific explainability regulations and proposed a conceptual framework for an explainable AV. Finally, we advocated for more interdisciplinary research around AV explainability. While uninhabited aerial vehicles (UAV) and autonomous underwater vehicles (AUV) were not discussed in this survey, we note that they can benefit from the provision of explanations and the recommendations we have provided in this paper. Hopes are high that AVs will bring many benefits to society. For these benefits to be realised, a further look at the recommendations provided in this paper for implementation purposes would be helpful.

\section*{Acknowledgment}
The authors would like to thank Ricardo Cannizzaro and Kala Allen for proof-reading the manuscript.

% if have a single appendix:
%\appendix[Proof of the Zonklar Equations]
% or
%\appendix  % for no appendix heading
% do not use \section anymore after \appendix, only \section*
% is possibly needed

% use appendices with more than one appendix
% then use \section to start each appendix
% you must declare a \section before using any
% \subsection or using \label (\appendices by itself
% starts a section numbered zero.)
%

% ============================================
%\appendices
%\section{Proof of the First Zonklar Equation}
%Appendix one text goes here %\cite{Roberg2010}.

% you can choose not to have a title for an appendix
% if you want by leaving the argument blank
%\section{}
%Appendix two text goes here.

% use section* for acknowledgement
%\section*{Acknowledgment}

%The authors would like to thank D. Root for the loan of the SWAP. The SWAP that can ONLY be usefull in Boulder...

% Can use something like this to put references on a page
% by themselves when using endfloat and the captionsoff option.
\ifCLASSOPTIONcaptionsoff
  \newpage
\fi

% trigger a \newpage just before the given reference
% number - used to balance the columns on the last page
% adjust value as needed - may need to be readjusted if
% the document is modified later
%\IEEEtriggeratref{8}
% The "triggered" command can be changed if desired:
%\IEEEtriggercmd{\enlargethispage{-5in}}

% ====== REFERENCE SECTION

%\begin{thebibliography}{1}

% IEEEabrv,

\bibliographystyle{IEEEtran}
\bibliography{IEEEabrv,Bibliography}
%\end{thebibliography}
% biography section
% 
% If you have an EPS/PDF photo (graphicx package needed) extra braces are
% needed around the contents of the optional argument to biography to prevent
% the LaTeX parser from getting confused when it sees the complicated
% \includegraphics command within an optional argument. (You could create
% your own custom macro containing the \includegraphics command to make things
% simpler here.)
%\begin{biography}[{\includegraphics[width=1in,height=1.25in,clip,keepaspectratio]{mshell}}]{Michael Shell}
% or if you just want to reserve a space for a photo:

% ==== SWITCH OFF the BIO for submission
% ==== SWITCH OFF the BIO for submission
\begin{IEEEbiography}[{\includegraphics[width=1in,height=1.25in,clip,keepaspectratio]{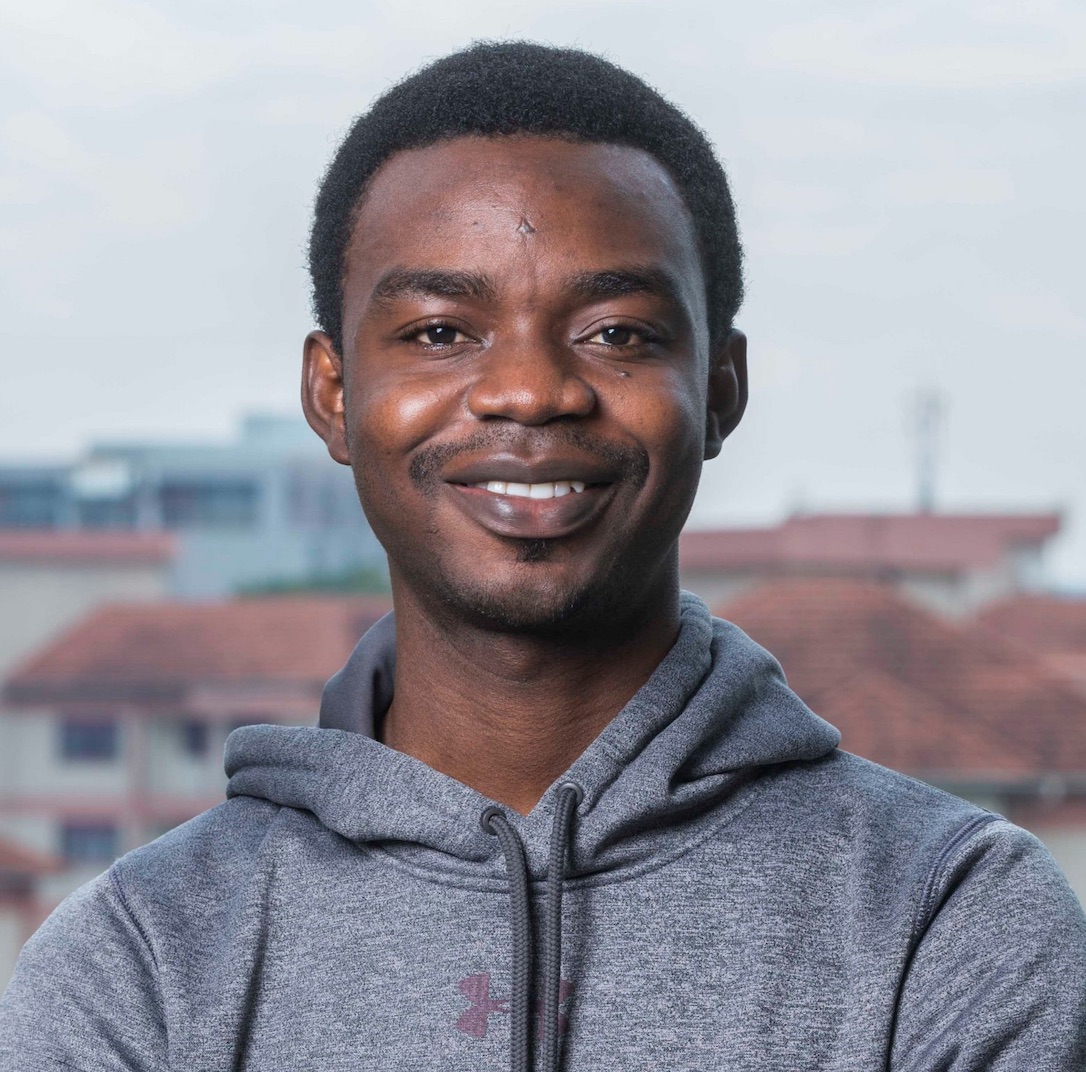}}]{Daniel Omeiza} received a bachelors degree in Computer Science from the University of Ilorin in 2015. He obtained a masters degree in Information Technology from Carnegie Mellon University in 2019. He is currently working towards a DPhil degree at the University of Oxford. His research interests primarily lie in explainability in autonomous driving. He is a student member of the IEEE.

\end{IEEEbiography}
\begin{IEEEbiography}[{\includegraphics[width=1in,height=1.25in,clip,keepaspectratio]{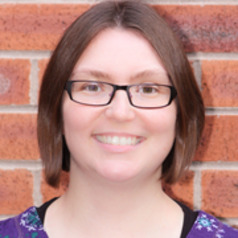}}]{Helena Webb} is a Transitional Assistant Professor in the School of Computer Science at the University of Nottingham. She is interested in the ways that users interact with technologies in different kinds of setting and how social action both shapes and is shaped by innovation. She works on projects that seek to identify mechanisms for the improved design, responsible development and effective regulation of technology. She has worked on projects relating to, amongst others, harmful content on social media, algorithm bias, resource scarcity in STEM education, human-robot dialogue, and responsible robotics. She is an interdisciplinary researcher and specialises in the application of qualitative research methods. She is also very interested in methodological innovation to combine detailed, granular analysis with larger scale computational work.

\end{IEEEbiography}
\begin{IEEEbiography}[{\includegraphics[width=1in,height=1.25in,clip,keepaspectratio]{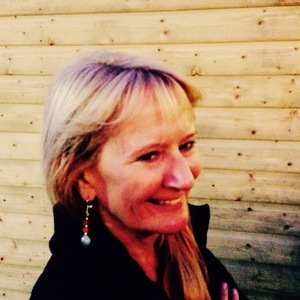}}]{Marina Jirotka} is Professor of Human-Centred Computing at the University of Oxford and Governing Body Fellow at St Cross College. She leads an interdisciplinary research group developing methods for building computing systems responsibly to support human, societal and environmental values. The team focuses on responsible innovation, in a range of ICT fields including robotics, AI, machine learning, quantum computing, social media and the digital economy. She is an EPSRC Established Career Fellow conducting a five-year investigation into Developing Responsible Robotics for the Digital Economy. She is Director of the newly established Responsible Technology Institute at Oxford, and she is co-director of the Observatory for Responsible Research and Innovation in ICT (ORBIT) which provides RI services and training to ICT researchers and practitioners.
She obtained her BSc in Psychology and
Social Anthropology (Hons) from The University of London Goldsmiths College in 1985 and her Masters in Computing and Artificial Intelligence from the University of South Bank in 1987. Her doctorate in Computer Science, ``An Investigation into Contextual Approaches to Requirements Capture", was undertaken at the University of Oxford in 2000.
\end{IEEEbiography}

\begin{IEEEbiography}[{\includegraphics[width=1in,height=1.25in,clip,keepaspectratio]{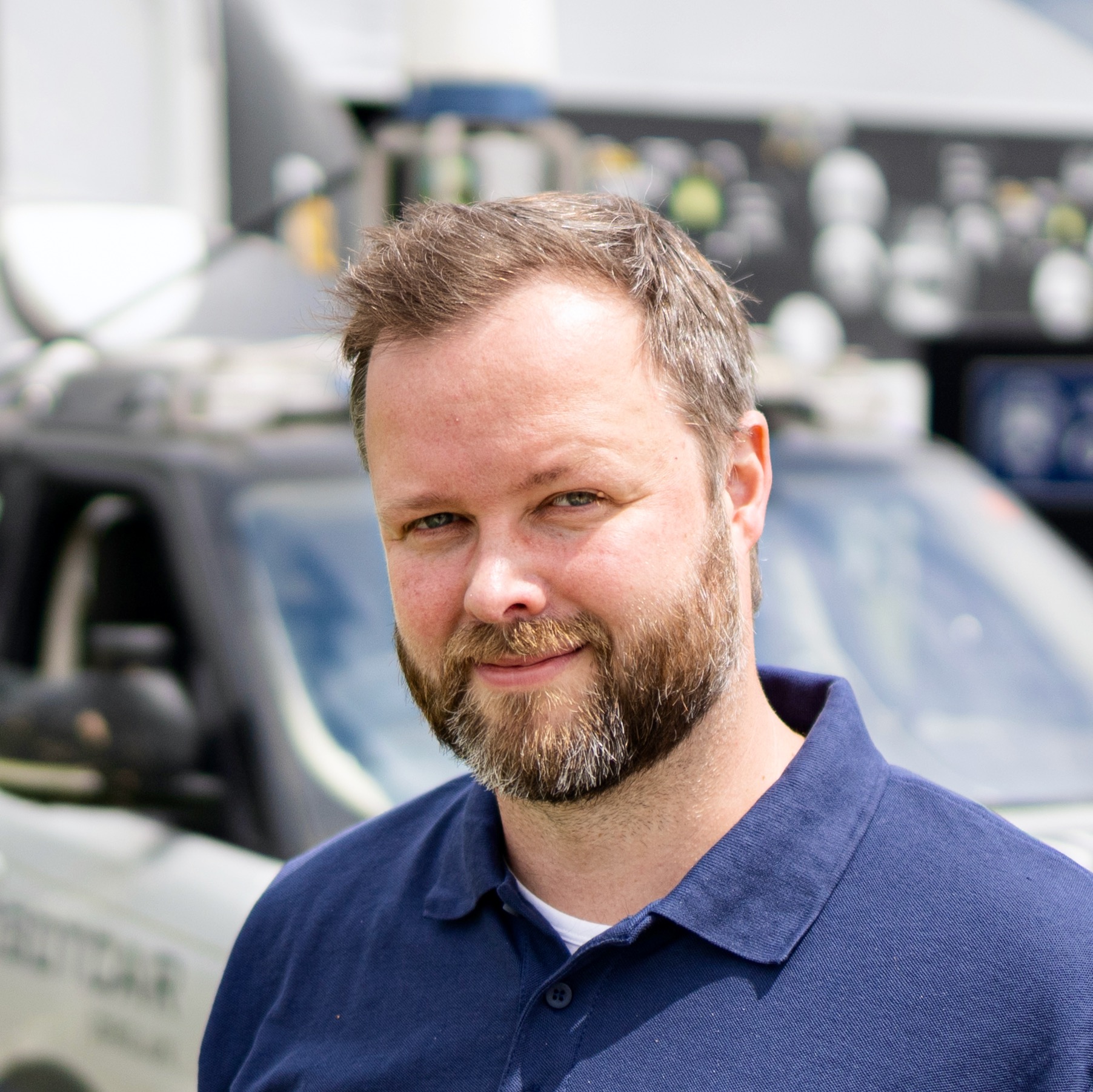}}]{Lars Kunze} is a Departmental Lecturer in Robotics in the Oxford Robotics Institute (ORI) and the Department of Engineering Science at the University of Oxford. In the ORI, he leads the Cognitive Robotics Group (CRG). Dr Kunze is also the Technical Lead at the Responsible Technology Institute (RTI), an international centre of excellence focused on responsible technology at Oxford University; and a Programme Fellow of the Assuring Autonomy International Programme (AAIP). He is a Stipendiary Lecturer in Computer Science at Christ Church and an Editor of both the Journal of Responsible Technology and the German Journal of Artificial Intelligence.
Before joining Oxford, he was a Research Fellow in the Intelligent Robotics Lab at Birmingham University. He studied Cognitive Science (BSc, 2006) and Computer Science (MSc, 2008) at the University of Osnabr\"uck, Germany, and partly at the University of Edinburgh,  UK. He received a PhD (Dr. rer. nat.) from the Technical University of Munich in 2014. 
\end{IEEEbiography}

%% if you will not have a photo at all:
%\begin{IEEEbiographynophoto}{Ignacio Ramos}
%(S'12) received the B.S. degree in electrical engineering from the University of Illinois at Chicago in 2009, and is currently working toward the Ph.D. degree at the University of Colorado at Boulder. From 2009 to 2011, he was with the Power and Electronic Systems Department at Raytheon IDS, Sudbury, MA. His research interests include high-efficiency microwave power amplifiers, microwave DC/DC converters, radar systems, and wireless power transmission.
%\end{IEEEbiographynophoto}

%% insert where needed to balance the two columns on the last page with
%% biographies
%%\newpage

%\begin{IEEEbiographynophoto}{Jane Doe}
%Biography text here.
%\end{IEEEbiographynophoto}
% ==== SWITCH OFF the BIO for submission
% ==== SWITCH OFF the BIO for submission

% You can push biographies down or up by placing
% a \vfill before or after them. The appropriate
% use of \vfill depends on what kind of text is
% on the last page and whether or not the columns
% are being equalized.

\vfill

% Can be used to pull up biographies so that the bottom of the last one
% is flush with the other column.
%\enlargethispage{-5in}

% that's all folks
\end{document}